\documentclass[%
 reprint,
superscriptaddress,
 amsmath,amssymb,
 aps,
]{revtex4-2}

\usepackage[utf8]{inputenc}
\usepackage[english]{babel}
\usepackage[T1]{fontenc}
\usepackage{amsmath}
\usepackage{tikz}
\usepackage{lipsum}
\usepackage{amsfonts}
\usepackage{amssymb}
\usepackage{hyperref}

\usepackage{tikz}
\usepackage{lipsum}
\usepackage{amsfonts}
\usepackage{amssymb}
\usepackage{graphicx}
\usepackage{multirow}
\usepackage{tabularx}
\usepackage{booktabs}
\usepackage{dcolumn}
\usepackage{bm}
\usepackage{subfig}
\usepackage{makecell}

\newtheorem{definition}{Definition}
\newtheorem{lemma}{Lemma}

\newtheorem{Theorem}{Theorem}

\begin{document}

\preprint{APS/123-QED}

\title{A Parameter-Efficient Quantum Anomaly Detection Method \\ on a Superconducting Quantum Processor}

\author{Maida Wang}
\affiliation{Centre for Computational Science, Department of Chemistry, University College London, UK}
\affiliation{Department of Computer Science, Tufts University, Medford, USA}

\author{Jinyang Jiang}
\affiliation{Guanghua School of Management, Peking University, Beijing, China}

\author{Peter V. Coveney}
\email{p.v.coveney@ucl.ac.uk}
\affiliation{Centre for Computational Science, Department of Chemistry, University College London, UK}
\affiliation{Advanced Research Computing Centre, University College London, UK}
\affiliation{Institute for Informatics, University of Amsterdam, The Netherlands}



\begin{abstract}
Quantum machine learning has gained attention for its potential to address computational challenges. However, whether those algorithms can effectively solve practical problems and outperform their classical counterparts, especially on current quantum hardware, remains a critical question. In this work, we propose a novel quantum machine learning method, called Parameter-Efficient Quantum Anomaly Detection (PEQAD), for practical image anomaly detection, which aims to achieve both parameter efficiency and superior accuracy compared to classical models.  Emulation results indicate that PEQAD demonstrates favourable recognition capabilities compared to classical baselines, achieving an average accuracy of over 90\% on benchmarks with significantly fewer trainable parameters. Theoretical analysis confirms that PEQAD has a comparable expressivity to classical counterparts while requiring only a fraction of the parameters.
Furthermore, we demonstrate the first implementation of a quantum anomaly detection method for general image datasets on a superconducting quantum processor. Specifically, we achieve an accuracy of over 80\% with only 16 parameters on the device, providing initial evidence of PEQAD's practical viability in the noisy intermediate-scale quantum era and highlighting its significant reduction in parameter requirements. 
\end{abstract}

\maketitle

\section{\label{sec:Intro}Introduction}
In recent years, research on quantum computing has seen rapid development across multiple areas~\cite{bernstein1993quantum,grover1996fast,shor1994algorithms}. For instance, research teams like that at the University of Science and Technology of China have demonstrated quantum advantage in boson sampling problems~\cite{zhong2020jiuzhang}. These achievements mark the advent of the Noisy Intermediate-Scale Quantum (NISQ) era~\cite{Preskill_2018}, where quantum computing systems can integrate dozens of noisy qubits. Despite the influence of noise, which limits these systems' capability for fault-tolerant universal quantum computation, they have demonstrated the potential to achieve performance beyond that of conventional computing in specific tasks. This has led current research to focus on leveraging quantum algorithms to achieve practical quantum advantage, with quantum machine learning being one of the most promising candidates and an active research hotspot.

Quantum machine learning (QML) aims to leverage quantum computing principles to enhance classical machine learning algorithms and address complex computational problems~\cite{biamonte2017quantum}. Recent advancements in quantum algorithms and hardware have enabled the exploration of quantum-enhanced models for a wide range of applications, such as quantum autoencoders for data compression, quantum Boltzmann machines for generative modelling, and quantum kernel methods for pattern recognition~\cite{23romero2021variational, 28huang2021power}. 

However, applying QML to datasets with complex and diverse characteristics remains challenging~\cite{kubler2021inductive}, particularly in practical scenarios involving noisy and limited quantum devices. Therefore, there is an urgent need for more examples demonstrating the practical feasibility of QML on widely recognized classical datasets, especially given the limitations of quantum hardware, such as the small number of available qubits. While QML shows potential, more work is needed to explore its capabilities within realistic constraints, using well-known datasets and conducting practical implementations on contemporary quantum devices~\cite{cerezo2022challenges}.

To date, many QML studies have focused on standard machine learning tasks, such as binary or multi-class classification. However, this paper aims to investigate the feasibility and efficacy of QML in anomaly detection. Fundamentally differing from binary classification which learns a boundary between two known classes, anomaly detection is typically framed as a one-class learning problem where the model is trained almost exclusively on 'normal' data. The objective is to learn a compact description of the normal class, such that any data that deviates significantly from this learned description is identified as an anomaly. Anomaly detection is a crucial machine learning field that seeks to identify unusual patterns or outliers in data, which can have significant implications for applications such as machinery monitoring, system intrusion detection, financial fraud prevention, and medical diagnostics~\cite{lavin2015evaluating3, garcia2009anomaly4, phua2010comprehensive5, schlegl2017unsupervised6}. This importance has motivated the development of numerous effective anomaly detection algorithms~\cite{an2015variational, 2018Deep15, YU2023, engelsma2019generalizing}. While various models have been proposed for anomaly detection on conventional computing platforms, the application of quantum machine learning to this task remains relatively unexplored. Only a handful of recent models attempt to use QML for anomaly detection \cite{park2023variational,belis2024quantum}, and existing algorithms have often struggled to achieve the performance levels of conventional machine learning methods, particularly on practical datasets~\cite{cerezo2022challenges}. Indeed, it has been demonstrated that a number of prevalent quantum models, as documented in the literature, exhibit comparable or inferior performance to classical machine learning on both classical and quantum datasets, even when utilizing emulators \cite{huang2021power}. This highlights the challenge of developing a QML algorithm that consistently outperforms classical methods for practical tasks.

Recent works, such as the kernel-based method applied to high-energy physics data \cite{belis2024quantum}, have explored the potential of QML in this domain. Our work complements these efforts but is distinct in its primary focus.

We concentrate on three key aspects: First, we aim to develop a novel quantum machine learning method for anomaly detection, striving to achieve competitive or even superior performance compared to classical methods in accuracy. Second, we explore the expressivity of PEQAD and conclude that PEQAD may have a larger learning space even with significantly fewer parameters. Thus, we intentionally constrained the number of learnable parameters in PEQAD to a mere fraction, often just a few per cent, of those used in typical classical deep learning models to investigate the potential parameter efficiency better. We also investigate the performance of PEQAD in scenarios with limited training data, as many real-world anomaly detection applications do not have large amounts of labelled data, requiring models that can learn effectively from small datasets.

Finally, we implement PEQAD on quantum hardware, making it, to the best of our knowledge,  the first end-to-end quantum anomaly detection algorithm applied to general image datasets implemented on a quantum device. Unlike prior studies that employed established methods and focused on domain-specific datasets, such as those from high-energy physics, our study demonstrates the versatility and scalability of PEQAD for general image datasets, with an emphasis on parameter efficiency. This experimental implementation serves multiple purposes: it demonstrates the compatibility of PEQAD on real hardware architecture and highlights its potential parameter efficiency in real quantum environments.

Our paper is organized as follows. In Section \ref{sec:Related Work}, we provide an overview of related work, including recent developments in quantum machine learning, anomaly detection, and quantum anomaly detection approaches. Section \ref{sec:quantum computing} covers the preliminaries of quantum computing, introducing the basics of variational quantum circuits. In Section \ref{sec:Algorithm}, we describe the structure of our proposed quantum machine learning algorithm, PEQAD, including its training process and data encoding strategy. Section \ref{sec:Characterization of PEQAD's Expressiveness} presents a theoretical analysis of PEQAD’s expressivity, comparing it to classical deep neural networks. In Section \ref{sec:Implementation Result}, we report its noiseless emulation and implementation on a superconducting quantum hardware platform. Finally, Section \ref{sec:conclusion} concludes the paper with a summary and discussion of future research directions.

\section{\label{sec:Related Work}Related Work}

\textbf{Quantum Machine Learning.}
QML has undergone rapid development, driven by advancements in quantum algorithms and hardware, allowing for the exploration of quantum-enhanced models across various applications. For instance, quantum autoencoders have been used for data compression and noise reduction in quantum systems, demonstrating the utility of QML in efficiently encoding quantum data with minimal qubit usage \cite{romero2017quantum16, 17lamata2018quantum, 18ding2019experimental}. Quantum Boltzmann machines have been proposed for quantum-state tomography and generative modelling, where they can learn complex quantum distributions \cite{19kieferova2017tomography, 20jain2020quantum}. Quantum generative adversarial learning has shown promise in learning and generating quantum states, with some successful applications in emulating quantum systems and verifying entanglement properties \cite{21dallaire2018quantum, 222018Quantum, 23romero2021variational, 24zeng2019learning}. Additionally, quantum kernel methods have been explored for pattern recognition tasks, leveraging quantum computing to improve the representational power of kernels for supervised learning \cite{25rebentrost2014quantum, 27mengoni2019kernel, 28huang2021power}.

While QML has shown considerable potential in addressing challenges involving quantum data, such as improving the efficiency of learning quantum systems or emulating complex quantum states \cite{huang2022quantum, cotler2021revisiting, chen2021hierarchy}, its practical application to datasets with complex and diverse characteristics remains challenging \cite{kubler2021inductive}. For instance, a recent experiment conducted by a collaboration between Caltech and Google demonstrated that a 40-qubit quantum computer could learn from significantly fewer samples compared to a conventional computer \cite{huang2022quantum}. Although this experiment was carefully controlled and not a practical case, it provide an encouraging glimpse into the potential of quantum machine learning.  Therefore, it remains essential to present more compelling and practical examples that showcase the strong performance of quantum machine learning on widely recognized and classical tasks. 

\textbf{Anomaly Detection.}
To address the unique challenges of one-class learning—where the goal is not to separate two known data classes but rather to identify anomalous patterns or outliers in data and has applications in monitoring machinery, detecting system intrusions, identifying financial fraud, and aiding medical diagnoses \cite{lavin2015evaluating3, garcia2009anomaly4, phua2010comprehensive5, schlegl2017unsupervised6}. Traditional anomaly detection algorithms include Parzen's density estimate \cite{parzen1962estimation8}, One-Class Support Vector Machine (OC-SVM) \cite{scholkopf2001estimating7}, and Support Vector Data Description (SVDD) \cite{39tax2004support}. With the progress of deep learning, deep AD algorithms such as AutoEncoders \cite{an2015variational}, Deep SVDD (DSVDD) \cite{2018Deep15}, Deep Convolutional AutoEncoder (DCAE) \cite{YU2023}, and one class generative adversarial network (OC-GAN) \cite{engelsma2019generalizing} have emerged, leveraging advanced representation learning techniques.

\textbf{Quantum Anomaly Detection.}
Applying QML to anomaly detection tasks remains relatively unexplored. Only a few recent models, such as variational quantum one-class classifier (VQOCC) \cite{park2023variational} and a kernel-based quantum anomaly detection method applied to the latent space of proton collision events at the LHC \cite{belis2024quantum}, have been proposed to utilize QML for anomaly detection. The former does not explore hardware implementation, whereas the latter achieves implementation but uses a quantum kernel method tailored to high-energy physics datasets,
with 300-dimensional input reduced to 4, 8, or
16 dimensions via autoencoders. In contrast, our work introduces PEQAD, a novel QML algorithm designed explicitly for anomaly detection on NISQ hardware, targeting general image datasets with original input sizes of 784 dimensions, reduced to 16 dimensions for hardware implementation. By emphasizing parameter efficiency and providing theoretical analysis, our study highlights the versatility of PEQAD in addressing general image anomaly detection tasks, making it more readily extendable to other tasks and larger scales.
In numerical emulations, PEQAD consistently outperforms classical and existing quantum anomaly detection methods on benchmark datasets with fewer parameters. Unlike previous works that mainly focus on emulation, we extended our study to real quantum hardware, where PEQAD achieved comparable performance with significantly fewer parameters. This aligns with our theoretical analysis, which suggests that PEQAD possesses a larger expressive capacity, even with only a fraction of the parameters used in classical models. Our hardware experiments further demonstrate the feasibility and robustness of PEQAD, showing its effectiveness in noisy quantum environments while maintaining low parameter requirements. These features make PEQAD particularly suited for small-scale quantum hardware and for addressing complex, high-dimensional anomaly detection tasks.

Our motivation for applying a quantum approach to anomaly detection is rooted in the task's central challenge: constructing a compact boundary around 'normal' data in a feature space. We propose that parameterized quantum circuits are particularly suited for this purpose. By mapping data into a vast Hilbert space, these circuits can generate rich feature representations ideal for defining a minimal hypersphere. We hypothesize that through quantum superposition and entanglement, such circuits can identify complex non-linear data correlations with significantly fewer parameters than classical counterparts. This parameter efficiency could enable the formation of a hypersphere with a sharper decision boundary, enhancing the model's sensitivity to subtle, unseen anomalies.

\section{\label{sec:quantum computing} Quantum Computing}

Quantum computing operates on principles rooted in quantum mechanics, utilizing quantum bits, or qubits, as the fundamental units of information. These qubits exist in a superposition of two basis states, commonly referred to as \( |0\rangle \) and \( |1\rangle \). A qubit state \( \psi \) can be expressed as \( \psi = \alpha|0\rangle + \beta|1\rangle \), where \( \alpha \) and \( \beta \) are complex amplitudes satisfying \( |\alpha|^2 + |\beta|^2 = 1 \). Measurement of a qubit causes the collapse of its superposition, with \( |\alpha|^2 \) and \( |\beta|^2 \) representing the probabilities of observing \( |0\rangle \) and \( |1\rangle \), respectively.

\begin{figure}[!tbp]
    \centering
    \includegraphics[width=0.45\textwidth]{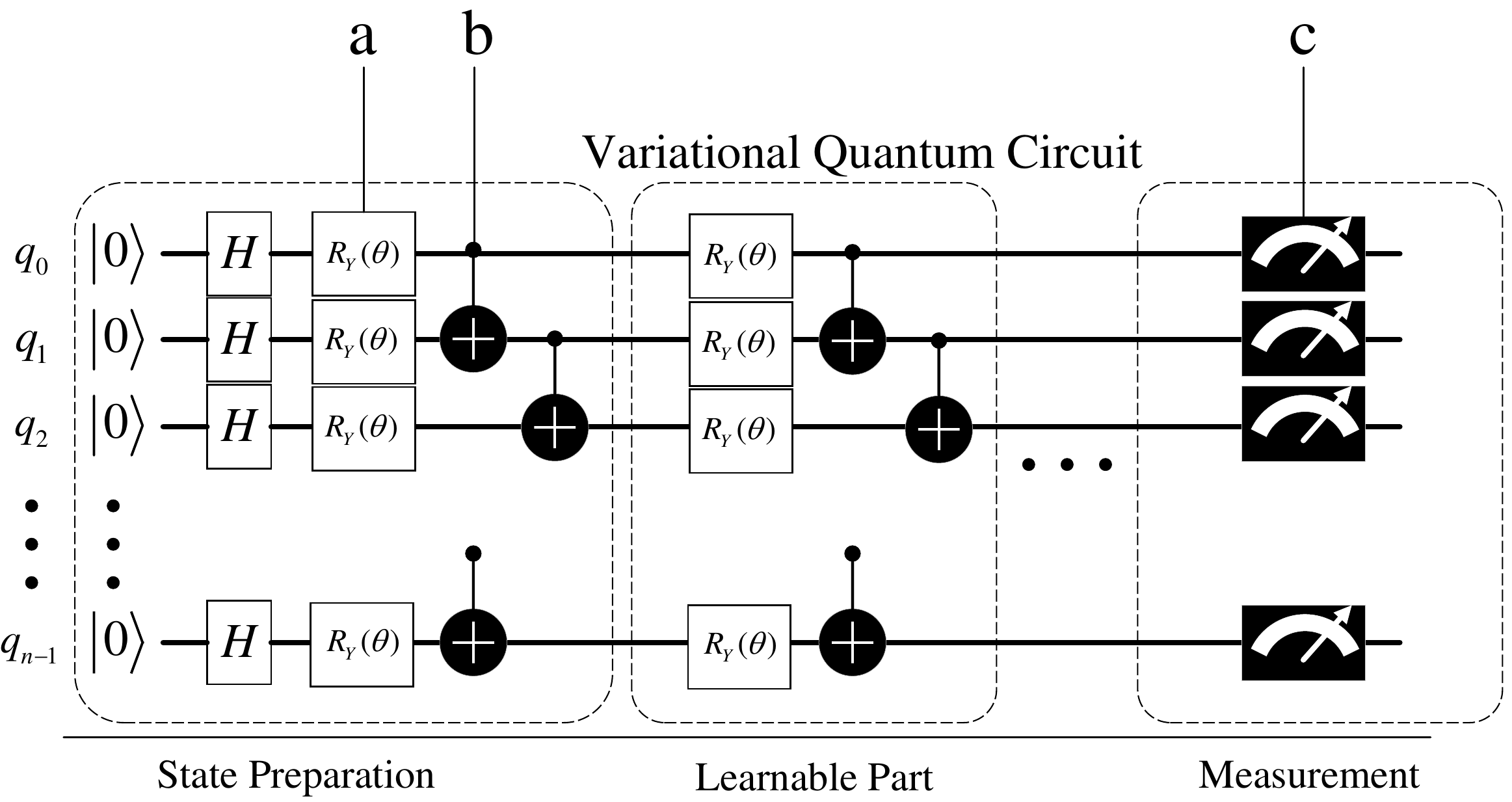}
    \caption{Diagram of the variational quantum circuit. (a) The first set of parameter-containing gates is used to embed classical features into the quantum circuit, and the rest of them are trainable gates. (b) CNOT gates for entanglement. (c) The measurement part maps quantum states back to classical vectors.}
    \label{fig:vqc}
    \end{figure}

Extending this concept to \( n \) qubits results in a composite state \( |s\rangle = \sum_{i=0}^{2^n-1} \alpha_i|i\rangle \), where \( |i\rangle \) denotes the $i$-th computational basis states for \( n \) qubits. The complex amplitudes \( \{\alpha_i\} \) satisfy the normalization condition \( \sum_{i=0}^{2^n-1} |\alpha_i|^2 = 1 \), representing the probability distribution of the state.
Quantum circuits utilize quantum gates to manipulate qubits. As illustrated in Fig. \ref{fig:vqc}, a variational quantum circuit (VQC) typically comprises several quantum gates, some of which are parameterized. Quantum gates operate as unitary transformations. Thus, a general parametric single-qubit quantum gate can be represented by a unitary matrix as follows:
\begin{eqnarray}
U(\boldsymbol{\theta}) = \left[
\begin{array}{cc}
e^{i(\alpha + \beta)}\cos(\frac{\gamma}{2}) & -e^{i(\alpha - \beta)}\sin(\frac{\gamma}{2}) \\
e^{i(\alpha - \beta)}\sin(\frac{\gamma}{2}) & e^{i(\alpha + \beta)}\cos(\frac{\gamma}{2})
\end{array}
\right],
\end{eqnarray}
where $\boldsymbol{\theta}=(\alpha, \beta, \gamma)$ denotes
parameters of the gate. These gates, represented as matrices, form the building blocks of quantum circuits. They manipulate qubits to perform operations and computations within the quantum computing framework. Those quantum gates transform one quantum state \( |x\rangle \) into another output state \( |y\rangle \) as denoted by \( |y\rangle=U(\boldsymbol{\theta})|x\rangle \), where quantum gate parameters control the functional relationship. Increasing the number of trainable gates allows for more complicated models. Certain quantum gates, such as the CNOT gate, enable qubit entanglement, establishing dependencies between qubit states. When a CNOT gate is applied, the state of the ``target'' qubit \( |t\rangle \) is flipped based on the state of the ``control'' qubit \( |c\rangle \), i.e.,
\begin{equation}
    \mathrm{CNOT} \, \left| c, t \right\rangle = \left| c, t \oplus c \right\rangle,
\end{equation}
where \( \oplus \) denotes the XOR operation. If the control qubit is in the state \( |1\rangle \), the target qubit is flipped; if it is \( |0\rangle \), the target remains unchanged. 
After executing the circuit, qubits are measured, yielding output probabilities for specific states. In VQC, this measurement process converts quantum states back into classical data. 

\begin{figure*}[!htbp]
    \centering
    \includegraphics[width=0.98\textwidth]{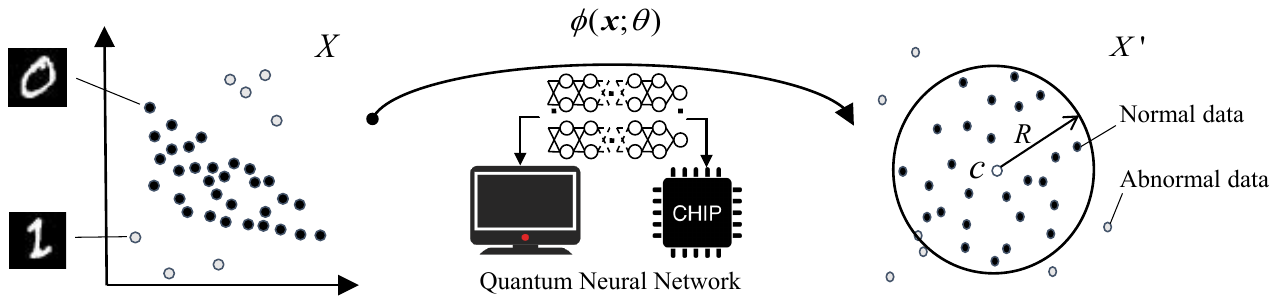}
    \caption{Conceptual diagram of Parameter-Efficient Quantum Anomaly Detection (PEQAD), which is a quantum-based anomaly detection algorithm that adopts QNN to map ``normal data'' into the hypersphere. Data points residing within the interior of the hypersphere will be labelled as ``normal data'', whereas those outside the hypersphere will be considered ``abnormal data''.}
    \label{fig:PEQAD}
    \end{figure*} 
    
The combination of trainable quantum gates and fixed ones is widely adopted in VQC, as well as in variational quantum eigensolvers (VQE), due to the convenience it offers in optimizing such variational structures. This variational approach, involving a mix of trainable and fixed components, has become a standard choice in many quantum neural networks (QNN), allowing them to extract complex data features and solve a wide range of problems \cite{romero2017quantum16,17lamata2018quantum,18ding2019experimental,28huang2021power}. Nevertheless, VQC is not without its limitations. Chief among these is the fact that the optimisation scale and depth are constrained by what is known as the barren plateaux problem, which refer to the exponentially vanishing gradients in variational quantum circuits, making optimization intractable \cite{mcclean2018barren}. However, as will be demonstrated in the following sections, our method is not subject to the aforementioned issue in real physical experiments.

\section{\label{sec:Algorithm}Methods}

This section details the design, architecture, and training methodology of our Parameter-Efficient Quantum Anomaly Detection (PEQAD), a novel QML algorithm specified for anomaly detection tasks.

\subsection{\label{sec:PEQAD}Modelling}

The main idea of PEQAD is illustrated by the conceptual diagram depicted in Fig. \ref{fig:PEQAD}. To be specific, PEQAD utilizes a QNN to transform raw data into a feature space, where the data points are supposed to be mapped into a hypersphere. The primary purpose of finding a hypersphere is to encompass a maximal amount of normal data while occupying a minimal volume. During the training stage of anomaly detection, normal data refers to the one class of data selected, while anomalies are considered unknown. Subsequently, the training process culminates in the acquisition of a learned hypersphere. During the testing process, data points whose QNN output falls outside the learned hypersphere are classified as anomalies, while those inside are considered normal.

Let us denote the given classical training data set as $X \subseteq \mathbb{R}^{d}$. $\phi \left  (  \cdot;\Theta  \right )$ is a QNN that encodes the real-valued vector $\boldsymbol x\in X$ into a corresponding quantum state $|x\rangle$ and then maps it back into the classical embedding set $X' \subseteq \mathbb{R}^{p}$ in the new feature space. The encoding process will be discussed in the next subsection. The QNN consists of a series of quantum gates, and some of them contain trainable parameters $\boldsymbol{\theta}
 = \left \{ \theta ^{1},\theta ^{2},\cdots,\theta ^{m} \right \}$. Thus, $\phi \left  ( \boldsymbol x;\boldsymbol{\theta} \right )\in X'$ is the mapped representation of $\boldsymbol x\in X$ given by the network structure $\phi $ with parameters $\boldsymbol{\theta}$. The hypersphere in the new feature space is characterized by a centre $\boldsymbol c\in X' $ and radius post-training $R>0$, which can be acquired in two ways during training \cite{2018Deep15}.  In our case, the radius is calculated as the mean distance between the centre of the hypersphere and the mapped points of the training dataset in the trained network's feature space.


For the training data set $X=\left \{\boldsymbol x_{1},\boldsymbol x_{2}..., \boldsymbol x_{n} \right \}$, the training of the QNN embedding model as well as the optimal hypersphere radius searching can be described as the following optimization:
\begin{align}
\min_{R,\boldsymbol{\theta}} \left\{ 
\begin{aligned}
  & R^2 +\frac{1}{vn}\sum_{i=1}^{n}\max\left \{ 0,\left \| \phi \left ( \boldsymbol x_{i};\boldsymbol{\theta} \right ) -\boldsymbol c\right \| ^{2}-R^{2}\right \} \\
  & + \frac{\alpha }{2}\sum_{j=1}^{m}\left \| \theta^{j} \right \|_{F}^{2} 
\end{aligned}
\right\} \label{fom:0}
\end{align}
where $\boldsymbol c = \frac{1}{n}\sum_{i=1}^{n}  \phi \left ( \boldsymbol x_{i};\boldsymbol{\theta} \right )$, the second term is the penalty for points falling outside the hypersphere boundary, and the last term is the regularization term to prevent overfitting on the training data. The trade-off between the sphere volume and violations of the boundary constraint is controlled by $v\in(0,1]$, while $\alpha>0$ is the regularization coefficient.

To solve Eq. (\ref{fom:0}) more effectively, a simplified formulation of PEQAD training can be defined in a manner analogous to that employed in \cite{2018Deep15}:
\begin{eqnarray}
\mathop{\min}  \limits_{\boldsymbol{\theta}} \ \frac{1}{n}\sum_{i=1}^{n}\left \| \phi \left  (\boldsymbol x_{i};\boldsymbol{\theta} \right )-\boldsymbol c \right \| ^{2} + \frac{\alpha }{2}\!\sum_{j=1}^{m}\left \| \theta^{j} \right \|_{F}^{2},
\label{fom:1}
\end{eqnarray}
where PEQAD shrinks the hypersphere by minimizing the collective distance of all data representations to the centre $\boldsymbol c$. Eq. (\ref{fom:1}) pushes the mapped data closer to the centre point, compelling the QNN to uncover shared patterns among elements in the dataset. Penalizing the overall distance across all data points, rather than excluding any outliers, supports the assumption that the bulk of the training data originates from one class. 
While testing, we define the anomaly score of the input data $\boldsymbol x\in X$ according to its calculated distance from the centre of the hypersphere, i.e.,
\begin{equation}
S(X)=\mathbf{1}\{ \left \| \phi \left  ( \boldsymbol x;\boldsymbol{\theta}^{*} \right ) -\boldsymbol c\right \|^{2}- R^2\geq a \},
\label{fom:2}
\end{equation}
where $R= \frac{1}{n}\sum_{i=1}^{n}\left \| \phi \left  (\boldsymbol x_{i};\boldsymbol{\theta} \right )-\boldsymbol c \right \| $, $\mathbf{1}\{\cdot\}$ denotes the indicator function, and $a>0$ is an adjustable threshold. When $S(X)=0$, PEQAD recognizes the test data as normal; otherwise, it is considered abnormal data. Next, we continue to describe the execution steps of PEQAD.

\subsection{\label{sec:Encodi}Data Encoding and \textit{Ansätze}}
To handle classical data, a QML algorithm needs a procedure that encodes them into quantum states \cite{giovannetti2008quantum,schuld2019quantum,lloyd2020quantum}. A straightforward way is to encode the data into the amplitudes of a state, i.e.,
\begin{equation}
x =  (x_1, x_2, \cdots, x_N)^{\top}\Rightarrow|x\rangle = \frac{1}{\|x\|} \sum_{i=1}^{N} x_i |i\rangle,
\end{equation}
where $N = 2^{n}$, and $n$ is the least number of the qubits required. This can be implemented on various quantum computing platforms \cite{silver2022quilt,patel2022optic} and is predominantly performed on emulators. The reason is that implementing amplitude encoding on actual quantum hardware requires a number of quantum gates that scale quadratically with the number of qubits, making it significantly more challenging for systems with a large number of qubits. To date, as far as we are aware, no successful implementation of this encoding method on hardware has been reported. However, because our algorithm requires fewer qubits, it becomes feasible to implement this process on hardware without introducing excessive noise.  With the amplitude encoding, we can greatly reduce the qubits used for data encoding to a logarithmic number $n=\lceil\log(D_c)\rceil$, where $D_c$ represents the dimensionality of classical data. 

We acknowledge that the scalability of preparing an arbitrary quantum state via amplitude encoding poses a significant challenge for near-term algorithms. This issue is particularly pronounced as the number of required gates can scale exponentially with the number of qubits for a general state, making the procedure impractical for larger systems. In the field of quantum many-body physics, this state preparation problem is often tackled using methods inspired by Matrix Product States (MPS), which can efficiently represent and prepare quantum states that possess a low-entanglement structure \cite{orus2014practical}. For such states, the gate complexity for preparation scales polynomially, often linearly, with the number of qubits, offering a viable path to scalability.

While our current work, as a proof-of-concept on a small-scale system, does not necessitate these advanced techniques, we recognize their critical importance. The successful extension of our method to larger quantum processors will likely rely on whether the feature vectors of pre-processed classical data correspond to quantum states amenable to efficient, MPS-based preparation. A systematic study of this structural property in common datasets remains a promising direction for future research.

\begin{figure}[!htbp]
    \centering
    \includegraphics[width=0.5\textwidth]{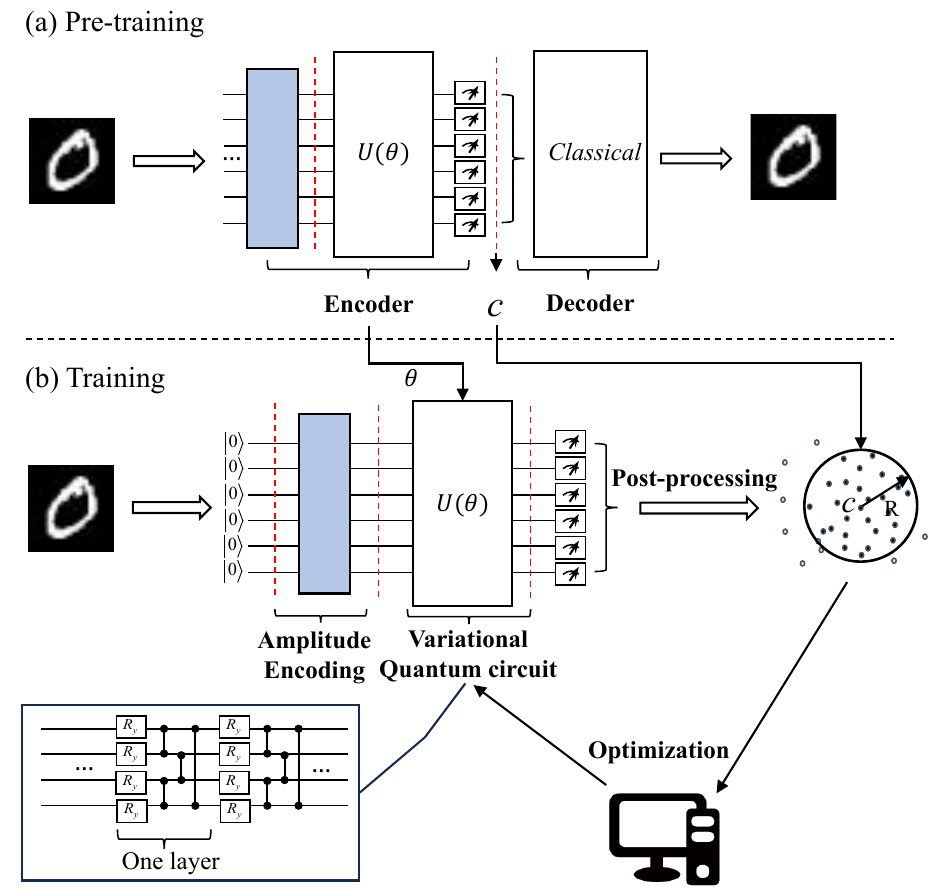}
    \caption{Training process of the PEQAD  algorithm. (a) The pre-training neural network of the method consists of two parts: a quantum encoding layer and a classical decoding layer, where the quantum encoding layer is consistent with the PEQAD network. After the quantum encoding layer, the centre of the hypersphere mapped by PEQAD can be obtained. (b) The PEQAD neural network is composed of an amplitude encoding part  (blue part), a VQC part, a measurement part and a PEQAD post-processing part. }
    \label{fig:PEQADall}
    \end{figure}

Following the quantum data encoding, a VQC is employed for optimization. Specifically, we adopt the \textit{ansätze} of the VQC shown in Fig.  \ref{fig:PEQADall}, which is derived from \cite{wang2023quantum}. The \textit{ansätze} comprise layers with parameterized single-qubit rotations $R_Y (\theta) = e^{-i\theta Y/2}$, followed by CNOT gates. These CNOT gates are arranged alternately between adjacent qubits. 
By employing these \textit{ansätze}, we can reduce the required hardware scale, which has several positive impacts. First, it helps mitigate the risk of barren plateaux \cite{mcclean2018barren}, thereby facilitating the optimization process. Second, it helps us use fewer learnable parameters, as the number of quantum gates required to train scales linearly with increased qubits when assuming a fixed number of quantum circuit layers.

\subsection{\label{sec:Train}Training Process}

\textbf{Pre-training:} As depicted in Fig.  \ref{fig:PEQADall} (a), the pre-training stage employs an encoder-decoder structure. The model for pre-training consists of a quantum encoding layer and a classical decoding layer, where the quantum one is synchronized with the PEQAD's network. During pre-training, the encoder maps the input data $\boldsymbol{x}$ to an embedding $\boldsymbol{x}'$, while the decoder aims to recover the original $\boldsymbol{x}$ from $\boldsymbol{x}'$. The parameters of the encoder are adjusted to minimize reconstruction errors by parameter shift rules \cite{wierichs2022general}. The centre of the hypersphere $\boldsymbol{c}$ is also initialized by averaging all the embeddings in $X'$. 
Both model parameters and a preliminary hypersphere centre are warmed up in this phase, accelerating the subsequent training stage.

\textbf{Training:} From this step onward, the PEQAD neural network is used for training and testing. Its structural details are illustrated in Fig. \ref{fig:PEQADall}(b). The network comprises an amplitude encoding part, a VQC part, a measurement part (represented by the dashboard), and a PEQAD post-processing part. The training stage mainly involves fine-tuning the PEQAD network. The optimization defined in Eq. (\ref{fom:1}) is utilized to minimize the volume of the hypersphere. Meanwhile, the hypersphere effectively encapsulates normal data patterns by reducing the average distance of all embeddings to the centre. The ratio of the time consumption between pre-training and training is found to be approximately $1.025:1$.

\section{\label{sec:Characterization of PEQAD's Expressiveness}Characterization of PEQAD's Expressivity}

To provide a theoretical basis and gain a better understanding of its performance, we analyze the expressivity of the PEQAD model. Expressivity is a crucial concept in quantum algorithms and has been extensively studied to understand the underlying theoretical foundations of effective machine learning models. Recent studies \cite[e.g.,][]{du2022efficient, holmes2022connecting} also focus on the expressivity of variational quantum algorithms (VQA). Understanding the theoretical expressivity of PEQAD is essential for gaining insights into its capabilities and potential advantages. We aim to reveal how PEQAD's expressivity influences its performance in tasks that may not conform to independently and identically distributed data rather than deriving generalization error bounds under such conditions.  In this section, we leverage the covering number \cite{vapnik2013nature}, a powerful concept from statistical learning theory, to investigate the expressive power of PEQAD. This analysis examines how the complexity of the hypothesis space influences PEQAD's ability to capture intricate data relationships. We target comparing PEQAD with other classical or quantum anomaly detection algorithms from different perspectives. By integrating empirical results with theoretical insights, we can have a comprehensive assessment of PEQAD's efficacy and its potential applications in QML. 

We begin by describing the process of the VQAs, which combine a $q$-qubit quantum circuit with a classical optimizer. During the training, the trainable parameters of the \textit{ansätze} are optimized iteratively based on the quantum circuit's outputs to minimize the objective function \( \mathcal{L}(\cdot,\cdot) \). 
The parameter update rule at iteration \( t \) is given by
\begin{eqnarray}
    \boldsymbol{\theta}^{(t+1)} &=& \boldsymbol{\theta}^{(t)} - \eta \left[ \frac{\partial \mathcal{L}(h(\boldsymbol{\theta}^{(t)}, O, \rho), c_1)}{\partial \boldsymbol{\theta}} \right],
\end{eqnarray}
where \( \eta \) is the learning rate, \( c_1 \in \mathbb{R} \) is the target label, and \( h(\boldsymbol{\theta}^{(t)}, O, \rho) \) is the quantum circuit's output. The gradient information $\partial \mathcal{L}(h(\boldsymbol{\theta}^{(t)}, O, \rho), c_1) / \partial \boldsymbol{\theta}$ can be obtained using the parameter shift rule or other methods.

In the context of anomaly detection, \( \rho \in \mathbb{C}^{d^q \times d^q} \) represents the q-qubit input quantum state, \( O \in \mathbb{C}^{d^q \times d^q} \) is the quantum observable, and \({U}(\boldsymbol{\theta}) = \prod_{l=1}^{P} {u}_l(\boldsymbol{\theta}) \in \mathbb{U}(d^q) \) denotes the \textit{ansätze}, with \( \boldsymbol{\theta} \in \Theta \) being the trainable parameters. We denote the \( l \)-th quantum gate acting on at most \( k \leq q \) qubits as \( {u}_l(\boldsymbol{\theta}) \in \mathbb{U}(d^k) \), and \( \mathbb{U}(d^q) \) is the unitary group of dimension \( d^q \). Generally, the \textit{ansätze} consists of \( P \) trainable gates, which correspond to the number of trainable parameters, i.e., \( \Theta \subseteq [0, 2\pi)^{P} \). The ideal output of the quantum circuit is given by:
\begin{equation}
    h(\boldsymbol{\theta}^{(t)}, O, \rho) = \mathrm{Tr}\left(U(\boldsymbol{\theta}^{(t)})^\dagger O U(\boldsymbol{\theta}^{(t)}) \rho\right).
\end{equation}

The definition of \( h(\boldsymbol{\theta}^{(t)}, O, \rho) \) can be generic. We now elucidate the connection between expressivity and model complexity in the context of PEQAD. The primary objective of PEQAD is to identify the best hypothesis \( h^{*}(\boldsymbol{\theta}, O, \rho) = \arg\min_{h(\boldsymbol{\theta}, O, \rho) \in \mathcal{H}} \mathcal{L}\left( h(\boldsymbol{\theta}, O, \rho), c_1 \right) \) that can adeptly approximate the target concept, where \( h(\boldsymbol{\theta}, O, \rho) \) and \(c_1\) correspond to \(\phi \left ( \boldsymbol x_{i};\boldsymbol{\theta} \right ) \) and \(\boldsymbol{c}\) in PEQAD modelling. We denote the hypothesis space of PEQAD as \(\mathcal{H}\), i.e.,
\begin{equation}
    \mathcal{H} = \left\{ (\mathrm{Tr}(U(\boldsymbol{\theta})^\dagger O U(\boldsymbol{\theta}) \rho) - \boldsymbol{c})^2 - R^2 \mid \boldsymbol{\theta} \in \Theta \right\},
\end{equation}
where \( R \) and \(\boldsymbol{c}\) retain the same definitions in previous sections.

\begin{figure}[!htbp]
 \centering
\includegraphics[width=0.48\textwidth]{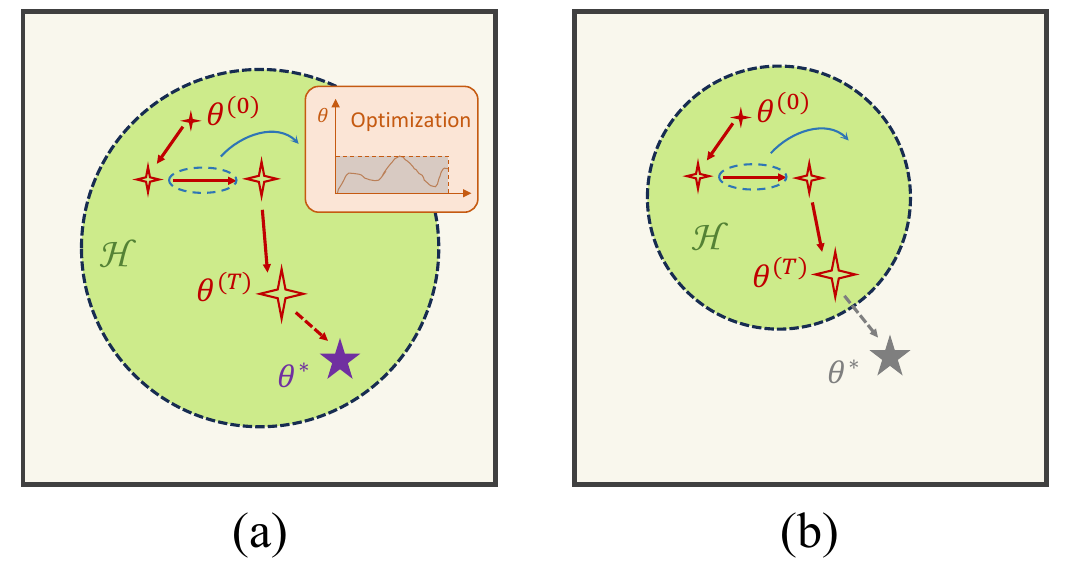}%
\caption{ Expressivity of PEQAD. If \(\mathcal{H}\)  is of moderate size and effectively encompasses the target concepts (purple solid star), the inferred hypothesis may closely approximate the target concept. }
\label{fig:SPACE}
\end{figure}

The geometric representation of PEQAD's expressivity is illustrated in Fig. \ref{fig:SPACE}, where the expressivity of the utilized \textit{ansätze} in PEQAD governs its hypothesis space denoted as \(\mathcal{H}\) (solid green ellipse).  
If \(\mathcal{H}\) is of a moderate size—meaning it is large enough to encompass the target concepts effectively but not excessively large—the inferred hypothesis may closely approximate the target concept. Conversely, if \(\mathcal{H}\) is too limited in complexity, there will be a substantial discrepancy between the estimated hypothesis and the target concept (grey solid star). Furthermore, if \(\mathcal{H}\) encompasses an excessively large space, optimization can become challenging.
Both overly high and low expressivities of PEQAD can result in suboptimal performance. Therefore, understanding and controlling the expressivity of PEQAD is a crucial task and requires an effective metric for evaluating the complexity of \(\mathcal{H}\). We utilize the covering number to establish an upper bound on the complexity of \(\mathcal{H}\), thereby quantifying the expressivity of PEQAD. This approach allows us to rigorously evaluate PEQAD's capacity to model complex distributions and highlight its theoretical capabilities.

\begin{definition}
The covering number \(N (U, \varepsilon, \| \cdot \|)\) represents the minimum cardinality of any subset \(V \subset U\) that effectively encompasses \(U\) with a scale of \(\varepsilon\) using the norm \(\| \cdot \|\), i.e., \(\sup_{A \in U} \min_{B \in V} \| A - B \| \leq \varepsilon\).
\end{definition}

The covering number $N (\mathcal{H}, \varepsilon, \|\cdot\|)$ signifies the minimal count of spherical balls, each with a radius of $\varepsilon$, necessary to cover a designated space, allowing for potential overlaps comprehensively.
$\varepsilon >0$ is a predefined small constant, which remains independent of any specific factors \cite{vapnik2013nature}. This convention is commonly embraced in the machine learning domain to assess model capacity in diverse learning scenarios \cite{mohri2018foundations}. The subsequent theorem provides the upper bound  of $N (\mathcal{H}, \varepsilon, \|\cdot\|)$ for PEQAD.


\begin{Theorem}\label{th}
For \(\varepsilon\in (0,\frac{1}{10})\), the covering number of the hypothesis space of PEQAD can be represented as follows:
\begin{equation}
N (\mathcal{H}, \varepsilon, \|\cdot\|) \leq \left (\frac{7P  (2\|O\| + 2\|\boldsymbol{c}\|)\|O\|}{\varepsilon}\right)^{q^{2m}P},\
\end{equation}
where \(\mathcal{H}\) denotes the function space represented by PEQAD, \(N (\mathcal{H}, \varepsilon, \|\cdot\|)\) represents its covering number, and \(\boldsymbol{c}\) is the centre of the hypersphere.
\end{Theorem} 

Theorem \ref{th} implies that the primary factor influencing the complexity of \(\mathcal{H}\) is the quantum gates incorporated into \({U}(\boldsymbol{\theta})\). This is evidenced by the fact that the term \(q^{2m} P\) contributes to an exponential increase in the complexity \(N(\mathcal{H}, \epsilon, \| \cdot \|)\). In contrast, the operator norm \(\|O\|\) and the centre \(\|\boldsymbol{c}\|\) affect the complexity in a polynomial manner.

\begin{Theorem}\label{th2}
For \(\varepsilon \in (0, \frac{1}{10})\), the supremum of the covering number of the hypothesis space of PEQAD is bounded as follows:
\begin{align}
    \left( \frac{7P \|O\|}{\varepsilon} \right)^{q^{2m}P} 
    &\leq \sup N(\mathcal{H}, \varepsilon, \|\cdot\|) \\
    &\leq \left( \frac{7P (2\|O\| + 2\|\boldsymbol{c}\|)\|O\|}{\varepsilon} \right)^{q^{2m}P}.
\end{align}

\end{Theorem}

Theorem \ref{th2} establishes that the supreme expressive capacity of our method is strictly greater than that of normal VQC-based anomaly detection models without post-processing like VQOCC. The lower bound shows that PEQAD can express at least the same function space as VQOCC, while the upper bound highlights its ability to capture a broader range of patterns due to its more flexible structure. This increased expressivity is crucial for more effective anomaly detection, particularly in complex real-world scenarios. PEQAD’s ability to model richer data distributions may offer an advantage over simpler VQC methods.

\begin{Theorem}\label{th3}
For a local depolarizing noise channel with depolarization rate \(p\), the expressivity of PEQAD on noisy quantum devices is bounded by:
\begin{small}
\begin{equation}
N(\mathcal{H},\varepsilon,\|\cdot\|) \leq (1-p)^{N_g} \left(\frac{7P(2\|O\|+2\|\boldsymbol{c}\|)\|O\|}{\varepsilon}\right)^{q^{2m}P},
\end{equation}
\end{small}
where \({N_g}\) is the number of gates affected by noise.

\end{Theorem}

Theorem \ref{th3} illustrates how noise, specifically modelled by the factor \((1 - p)^{N_g}\), reduces the expressivity of PEQAD. Here, \(p\) represents the noise rate in a depolarizing noise channel, and \(N\) is the number of quantum gates affected by the noise. As the noise rate \(p\) increases (i.e., as \(p\) approaches 1), the noise factor \((1 - p)^{N_g}\) approaches zero, which implies a significant reduction in the effective hypothesis space. This shrinking of the hypothesis space limits the PEQAD model's ability to learn and represent complex patterns, particularly in noisy quantum environments. Nevertheless, given that PEQAD encompasses a more expansive expressivity than other VQC anomaly detection algorithms with an equivalent number of parameters, it is more resilient to noise. The theorems above establish a theoretical upper bound on PEQAD's expressivity under different circumstances, demonstrating its capability to model functions in high-dimensional spaces. 
Detailed mathematical foundations supporting these findings are provided in the appendices. 

\textbf{Limitation.} Since our work mainly considers anomaly detection tasks, the training set typically lacks outliers in general anomaly detection scenarios, while the test set includes outlier data. Consequently, statistical analysis of these sets is limited to analyzing the model's complexity, which is the focus of this subsection. That is, we are unable to exhibit the superiority of PEQAD further using an expression for the generalization error of the same distribution. However, integrating theoretical and experimental results allows us to demonstrate that the theoretical analysis aligns well with the emulation and experimental outcomes. This consistency provides deeper insights into PEQAD's superior expressive capacity, which enables it to effectively capture complex patterns, leading to better performance in practical anomaly detection tasks. The detailed discussion in the appendices elaborates on how the enhanced expressivity of PEQAD could further help harness the power of QML to address anomaly detection challenges.

To compare the expressivity of PEQAD with classical algorithms such as DSVDD, we also provide the theoretical covering number bound for classical models. This bound originally comes from Bartlett et al.'s work \cite{NIPS2017_b22b257a}, and we further derived its form to focus on the impact of parameters on the model's expressivity. These derivations are presented in the supplementary materials, and the resulting bound is as follows:

\begin{Theorem}\label{th:classical}
For a fixed training set with data norm bounded by a constant \( t \), and for \(\varepsilon \in (0, \frac{1}{10})\), the covering number of the hypothesis space of DSVDD and other classical deep networks, denoted as \(\mathcal{H}_c\), can be represented as follows:
{\small
\begin{align}
\ln N (\mathcal{H}_c, \varepsilon, \|\cdot\|_2) &\leq \frac{t^2 \left( \ln(2) + 2 \ln(\max(m_j \times n_j)) \right)}{\varepsilon^2} \\
&\times \left( \prod_{j=1}^{L} m_j n_j \|W_j\|_{\infty}^2 \right) 
\left( \sum_{i=1}^{L} \left( \frac{\sqrt{m_i}}{\sqrt{n_i}} \right)^{2/3} \right)^3,
\end{align}
}where \( N (\mathcal{H}_c, \varepsilon, \|\cdot\|_2) \) represents the covering number of hypothesis space \(\mathcal{H}_c\) of a classical neural network, \(m_j\) and \(n_j\) are the dimensions of the weight matrices in each layer, \(W_j\) is the weight matrix of the \(j\)-th layer, and \( \|W_j\|_{\infty} \) represents the maximum absolute value of the elements in \(W_j\). The constant \( t \) is the norm of the data matrix, and \( L \) is the number of layers in the network.
\end{Theorem}

An important consideration is that increasing the expressive power of a QML algorithm generally increases the likelihood of encountering the barren plateaux problem. To examine this potential issue, we conduct a theoretical analysis to assess whether our algorithm is particularly prone to the barren plateaux.
We find that, for our PEQAD algorithm, assuming the unitaries in the circuit form a \(t\)-design with \(t \geq 2\), the algorithm remains susceptible to the barren plateaux phenomenon. However, the rate of gradient decay of the method is not accelerated compared to other variational quantum methods under similar conditions. We further analytically confirm that the PEQAD ansatz, when implemented on our specific hardware-efficient circuit, does not suffer from barren plateaus.
This indicates that PEQAD enhances expressivity without the BP problem in our case, maintaining optimization stability even for large-scale quantum circuits.

Based on the results above, we derive the following specific conclusion regarding the expressivity of PEQAD compared to classical models.

\textbf{Conclusion.}  PEQAD can achieve comparable or even superior performance using significantly fewer learning parameters, offering a potential advantage for scaling quantum machine learning in practical applications. We validate this observation both theoretically and experimentally. In the experimental section, we show that PEQAD, with only 16 parameters, achieves recognition performance equivalent to that of a classical neural network with approximately 400 parameters. In the supplementary theoretical analysis, we provide evidence suggesting that the expressivity of PEQAD, even with 16 parameters on quantum hardware, can reach a similar level to that of a classical neural network with hundreds of parameters. Additionally, we demonstrate that this enhanced expressivity algorithm does not have the BP phenomenon, further supporting the practical viability of PEQAD. We provide a detailed derivation of this result in the Appendix, where we compute the exact gradient statistics under our circuit design.

\section{\label{sec:Implementation Result}Implementation and Results}



\subsection{\label{sec:Results}Noiseless Emulation Results}
Considering that the vast majority of QML algorithms, including quantum anomaly detection approaches, are typically evaluated in emulated, noise-free environments, we also provide results in this way to establish a fair comparison. Specifically, we first implement our PEQAD on three widely recognized image datasets: MNIST \cite{lecun1998mnist}, FashionMNIST \cite{xiao2017fashion}, and CIFAR-10 \cite{krizhevsky2009learning}. For each dataset, data from a single class is used as the normal training set, while the other classes are treated as anomalies. Performance is assessed based on ten independently replicated emulations using the Area Under the Curve (AUC) metric, which is standard in anomaly detection \cite{perera2021one}. It represents the area under the Receiver Operating Characteristic curve. It serves as a standard metric to assess the model's ability to distinguish between normal and anomalous data points, ranging from 0.5 (random guess) to 1.0 (perfect classification). 
Given the computational overhead of quantum measurement, we constrain the size of each training set \( S_M \) by randomly sampling $300$ images from the original class, which contains normal images from class \( M\in\{0,\cdots,9\} \), respectively. 
Emulations are conducted using the PennyLane emulator integrated with a PyTorch-based Adam optimizer \cite{bergholm2018pennylane, 37kingma2014adam}, enabling efficient parameter optimization through auto-differentiation.
We use a batch size of $50$, a learning rate of $1\times10^{-3}$, and $150$ epochs for Adam to optimize the model. We vary the number of VQC parameters $P$ from $120$ to $280$ and that of qubits $Q$ from $10$ to $16$. If not specified, $P$ and $Q$ are set to $200$ and $10$ as default in the subsequent discussions.
We benchmark PEQAD against established anomaly detection methods such as DCAE \cite{YU2023}, OC-GAN \cite{YU2023}, LSP-CAE \cite{YU2023}, DSVDD \cite{2018Deep15}, and QHSVDD \cite{wang2023quantum} to evaluate its efficacy rigorously. These comparisons aim to uncover PEQAD's advantages over both classical and hybrid quantum approaches, offering insights into its potential applications.

\begin{table}[!htbp]
    \centering
    \caption{Comparison via emulation results. }
    \resizebox{0.95\linewidth}{!}{
 \begin{tabular}{ccccccc}
    \toprule
    Dataset & Network & Train size & $Q$ & $P$ & AUC\\
    \midrule
     \multirow{7}{*}{MNIST} &OC-GAN& all  & /& $>$6500 & 90.57\% \\
     &DCAE &all&	/&$>$6500	&89.96\% \\
     &DSVDD & all  & /& $>$6500 & 94.80\% \\
     &DSVDD& 300  & /& 6500 & 86.59\% \\
     &QHDSVDD &300&	16&6500	&88.24\% \\
     &DSVDD& 300  & /& 200 & 79.44\% \\
     &PEQAD (Ours) &300&	10&200	&92.26\% \\
     &PEQAD (Ours)&all&	10&200	&92.31\% \\
     \midrule
      \multirow{7}{*}{\makecell[l]{Fashion\\ MNIST}}	&OC-GAN & all  & /& $>$6500 & 86.93\% \\
      &DCAE &all&	/&$>$6500	&90.73\% \\
      &DSVDD & all  & /& $>$6500 & 88.50\% \\
     &DSVDD& 300  & /& 6500 & 88.20\% \\
     &QHDSVDD &300&	16&6500	&88.19\% \\
     &DSVDD& 300  & /& 200 & 81.59\% \\
     &PEQAD (Ours)&300&	10&200	&91.35\% \\
    &PEQAD (Ours)&all&	10&200	&91.81\% \\
        \midrule
      \multirow{6}{*}{CIFAR-10}	&OC-GAN & all  & /& $>$6500 & 65.63\% \\
     &LSP-CAE &all&	/&$>$6500	&67.63\% \\
      &DSVDD & all  & /& $>$6500 & 64.81\% \\
      &DSVDD& 300  & /& 6500 & 63.95\% \\
     &DSVDD& 300  & /& 200 & 57.25\% \\
     &PEQAD (Ours)&300&	10&200	&64.65\% \\
    &PEQAD (Ours)&all&	10&200	&63.31\% \\
    \bottomrule
 \end{tabular}}
\label{table: main results}
\end{table}

In Tab. \ref{table: main results}, we exhibit the overall AUC scores averaged among multiple independent emulations for each method mentioned above on three datasets, where different classes of data are selected as the normal input each time. We denote the number of trainable parameters in our PEQAD and other classical deep anomaly detection neural networks as $P$, while $Q$ stands for the number of qubits if applicable.

\begin{figure}[!tbp]
    \centering 
    \subfloat[]{
        \label{fig:AUCwithdepth}
        \includegraphics[trim=0cm 0cm 0cm 0cm, width=0.215\textwidth]{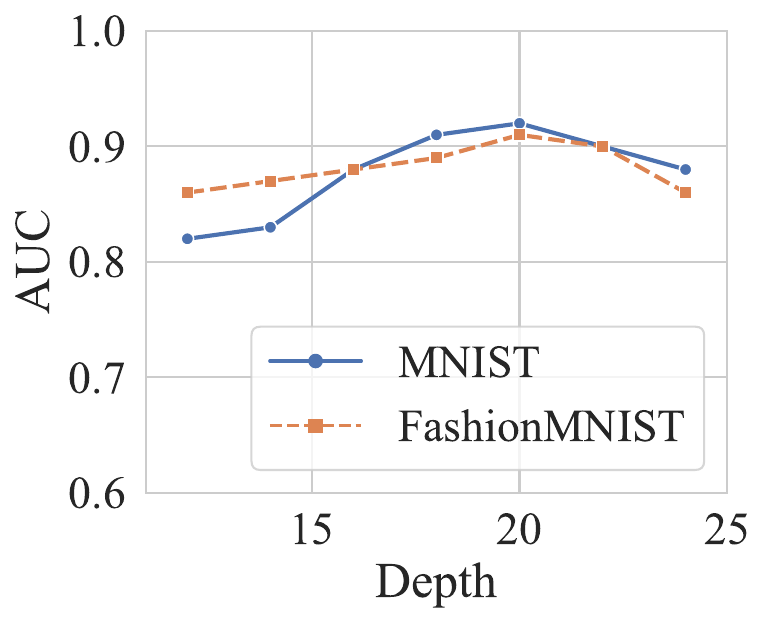}
    }
    \hspace{0.01\textwidth}
    \subfloat[]{
        \label{fig:AUCwithvqc}
        \includegraphics[trim=0cm 0cm 0cm 0cm, width=0.215\textwidth]{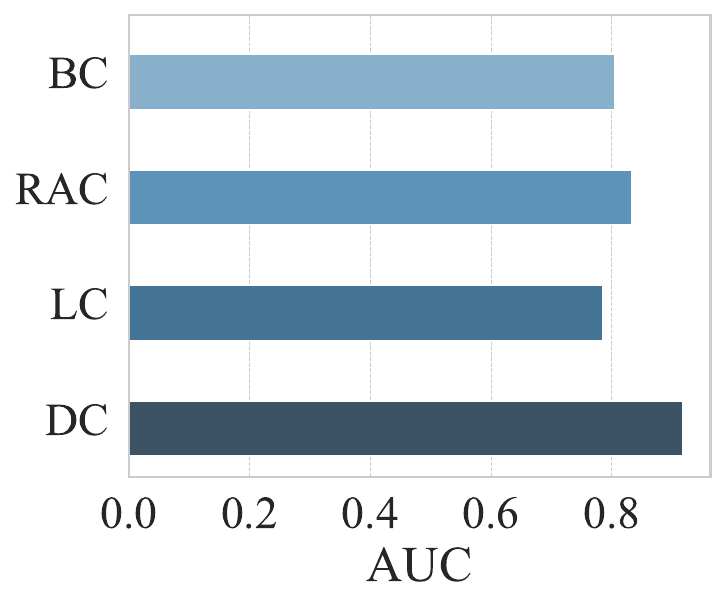}
    }
    \hspace{0.01\textwidth}
    \subfloat[]{
        \label{fig:AUCwithparameter}
        \includegraphics[trim=0cm 0cm 0cm 0cm, width=0.215\textwidth]{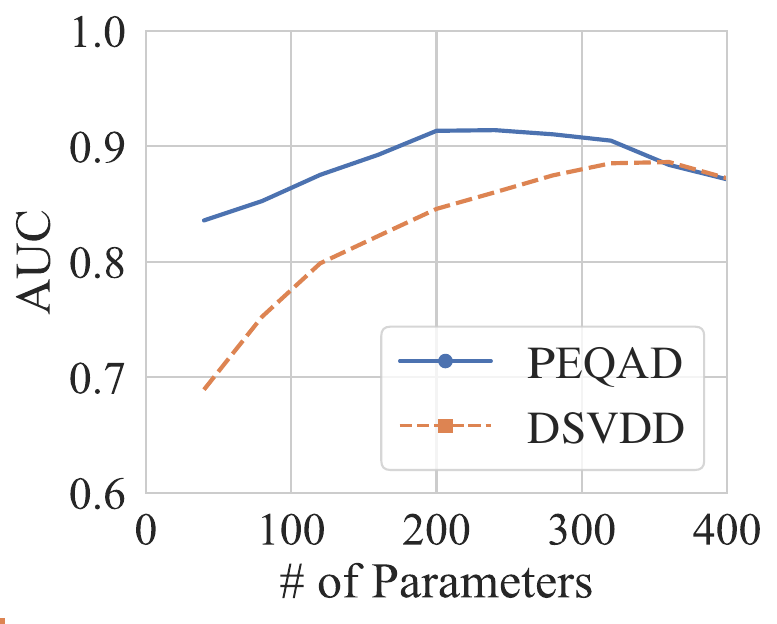}
    }
    \hspace{0.01\textwidth}
    \subfloat[]{
        \label{fig:AUCwithepoch}
        \includegraphics[trim=0cm 0cm 0cm 0cm, width=0.215\textwidth]{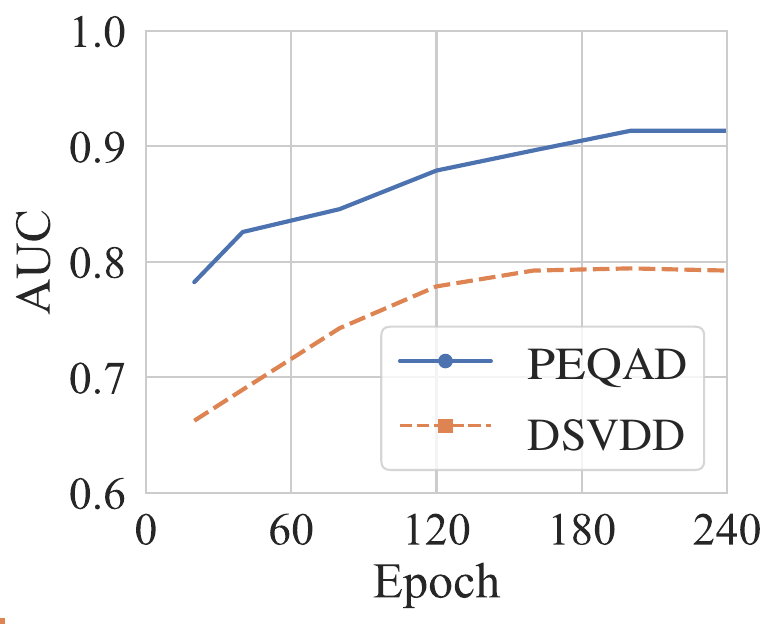}
    }
    \vspace{-0.2cm}
    \caption{Ablation study on the effects of VQC's depth, structure, number of parameters, and training epochs. Although the effect of the PEQAD and its conventional counterpart DSVDD varies with settings variation, PEQAD always outperforms DSVDD under the same conditions.}
    \label{fig:AUC}
    \vspace{-0.3cm}
\end{figure}

PEQAD consistently outperforms classical solutions across most evaluations, even when the latter employs significantly more parameters. For instance, on the MNIST dataset, PEQAD achieves an AUC of \( 92.26\% \) with only $200$ parameters and $300$ training data, surpassing DSVDD by $5.67\%$. This relative improvement widens to $12.82\%$ when both methods use an equal number of parameters.
Similar trends are observed on FashionMNIST, where PEQAD achieves an AUC of \( 91.35\% \) with $200$ parameters and $300$ data, outperforming DSVDD by $6.76\%$. Furthermore, PEQAD demonstrates superior performance to DSVDD on the FashionMNIST dataset when utilizing the complete training set. The MNIST and Cifar-10 datasets exhibit slightly inferior results compared to DSVDD, which has 6500 parameters. Nevertheless, it still exhibits considerably enhanced performance compared to DSVDD with 200 parameters. Compared to other deep neural networks like DCAE, which utilizes over $6500$ parameters to achieve an AUC of \( 89.96\% \) on MNIST, PEQAD demonstrates superior performance with significantly fewer computational resources. Even under more substantial training conditions, PEQAD maintains higher accuracy levels, showcasing its efficiency in anomaly detection tasks.
PEQAD also exhibits competitive performance with much lower consumption on the CIFAR-10 dataset, demonstrating its ability to handle complex data structures. The robust and scalable performance of PEQAD in our emulations indicates a potential for quantum advantage in anomaly detection tasks. However, further investigation and experimental verification are required to demonstrate this advantage conclusively. These emulation results demonstrate PEQAD's viability for practical applications and indicate that it is a perfect candidate for future experiments on NISQ devices, which will be discussed in the next section. To further investigate the effect of different configurations, we performed an ablation study as shown in Fig. \ref{fig:AUC}, which will be detailed in the following discussion.


\textbf{Depth of VQC.}  
We vary the depth $D$ of the VQC within specified ranges, and Fig. \ref{fig:AUCwithdepth} illustrates how the performance of PEQAD changes with different values of $D$. When $D$ ranges from $12$ to $24$, PEQAD consistently achieves satisfactory AUC scores (above $80\%$) on both datasets. Optimal results on both MNIST and FashionMNIST are observed at $D = 20$. However, increasing $D$ beyond $20$ leads to a decline in AUC, likely due to overfitting caused by excessive trainable parameters. Therefore, we default to setting  $D$ to $20$ for subsequent analyses.

\textbf{Different VQC Structures.} 
We conduct emulations using various VQC architectures: Real Amplitudes Circuit (RAC) \cite{sebastianelli2021circuit}, Bellman Circuit (BC) \cite{sebastianelli2021circuit}, Ladder-Like Circuit (LC) \cite{zeng2022multi}, and Dressed Quantum Circuit (DC) \cite{mari2020transfer}. The results are depicted in Fig. \ref{fig:AUCwithvqc}, demonstrating that varying VQC structures yield different performance outcomes. DC notably outperforms other structures, which is attributed to its alternating arrangement of entanglement gates across adjacent qubits, thereby enhancing the stability of circuit-wide entanglement and resulting in superior performance. Therefore, we use DC as the VQC structure in the emulations if not specifically stated.

\textbf{Parameters and Epochs.}
The number of parameters and epochs can influence the performance of our PEQAD solution and other anomaly detection methods. 
Fig. \ref{fig:AUCwithparameter} demonstrates that PEQAD achieves robust performance on the FashionMNIST dataset when the number of parameters varies from $100$ to $400$, showing higher stability compared to its classical counterpart, DSVDD. Additionally, Fig. \ref{fig:AUCwithepoch} illustrates that varying the number of training epochs from $40$ to $240$ consistently maintains PEQAD's high-performance level. However, beyond $200$ epochs, both PEQAD and classical DSVDD experience a decline in AUC, which may be attributed to overfitting. Notably, PEQAD exhibits superior and more stable performance under most settings than its classical counterpart.

\begin{figure}[!htbp]
\centering
\includegraphics[width=0.48\textwidth]{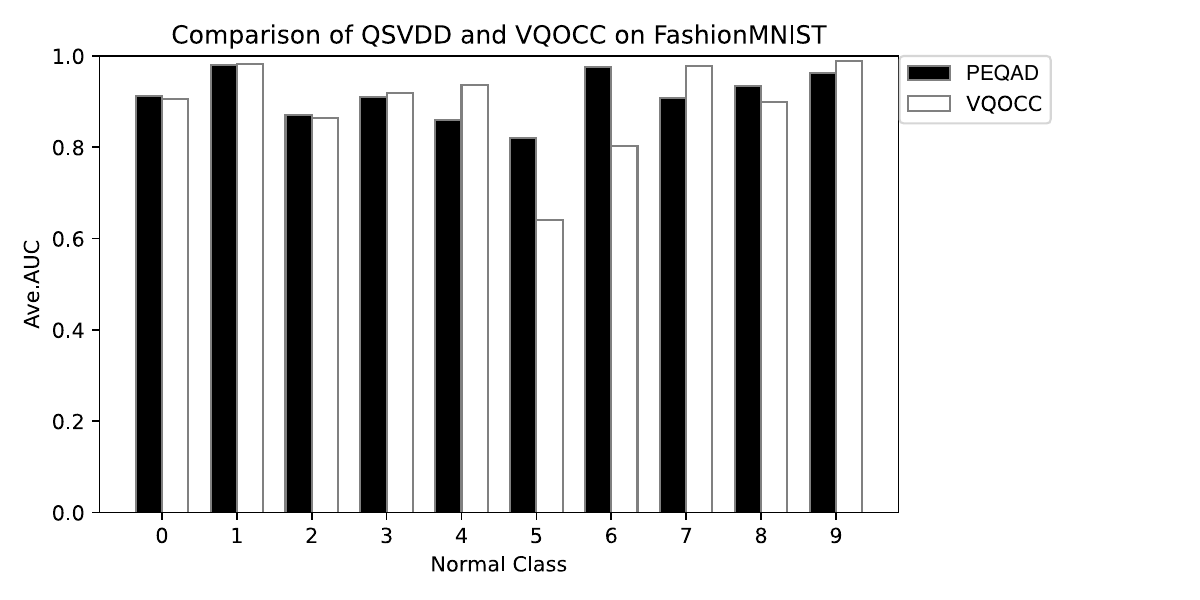}%
\caption{ Comparison of PEQAD and Variational Quantum One Class Classifier (VQOCC). }
\label{fig:vqocc}
\end{figure}

\textbf{PEQAD and other QML anomaly detection methods.}
As illustrated in Fig. \ref{fig:vqocc}, the comparison between PEQAD and another QML anomaly detection method, Variational Quantum One-Class Classifier \cite{park2023variational}, shows the performance difference within the quantum-based algorithm family on specific anomaly detection tasks. PEQAD consistently achieves results comparable to the variational quantum one-class classifier across most categories and significantly outperforms it in challenging scenarios where the variational quantum one-class classifier fails (e.g., Classes 5 and 6), demonstrating its superior robustness and reliability. Another promising quantum anomaly detection algorithm has recently been published, which uses kernel-based methods \cite{belis2024quantum}. However, as it has only been tested on specific high-energy datasets, a direct comparison with our work is not currently feasible. These results underscore the effectiveness of PEQAD and its suitability for a wider range of anomaly detection tasks. 

\subsection{\label{sec:EResults}Results on a Superconducting Quantum Processor}

\begin{figure}[!tbp]
\centering
\includegraphics[width=0.48\textwidth]{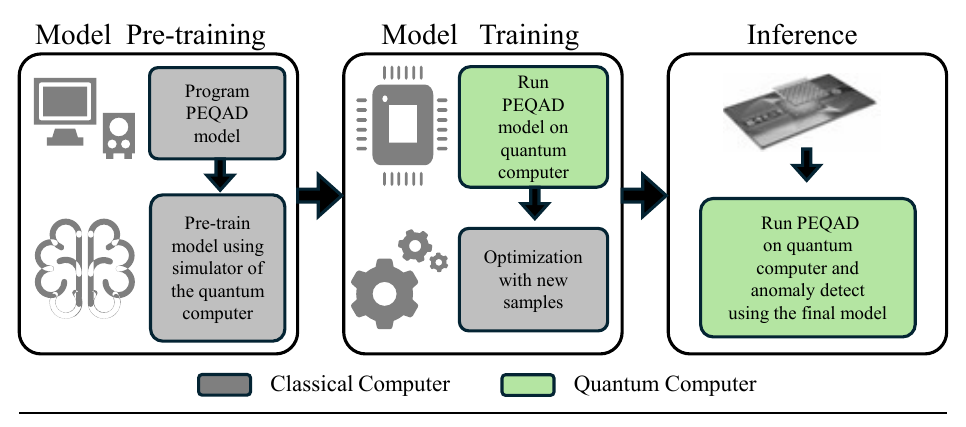}%
\caption{ Practical pipeline to address anomaly detection tasks on a quantum computer in the NISQ era. }
\label{fig:PEQADONQC}
\end{figure}

For the first time, we implement a quantum anomaly detection method for general image datasets on quantum hardware. To thoroughly evaluate the performance of PEQAD, we conducted experiments on a 12-qubit superconducting processor oneD12 provided by the Institute of Quantum Information and Quantum Technology Innovation, Chinese Academy of Sciences (CAS) \cite{oneD12_CAS}. To ensure fairness, we kept all experimental emulation settings, including datasets and training conditions, consistent across the hardware platform and emulators. Specifically, we used only four qubits for the hardware implementation, facilitating the realization of amplitude encoding. The fidelity of the device used in our experiments can be found on the official websites of the Institute of Quantum Information and Quantum Technology Innovation, CAS. For instance, the single-qubit gate fidelity for state 0 of the oneD12 chip we used averages at $0.9965$ (ranging from $0.9946$ to $0.9997$), while the CZ gate fidelity averages at $0.9634$ (with a range from $0.9354$ to $0.9885$). In order to provide a comprehensive demonstration of the performance and robustness of our algorithm in the presence of noise, and given the limited scale of the superconducting chip employed, we elected not to apply any noise reduction techniques or error mitigation methods to the implementation of our method on the quantum device.

During training on the quantum processor, post-processing was applied to the measurement data to map learned states onto the hypersphere, following Eq. (\ref{fom:1}). Once training was completed, the model was evaluated on the test set using Eq. (\ref{fom:2}) to determine if the data points were anomalies. The overall experimental pipeline for implementing PEQAD on quantum hardware in the NISQ era is illustrated in Fig. \ref{fig:PEQADONQC}. 

\begin{figure*}[!tbp]
\centering
\includegraphics[width=0.95\textwidth]{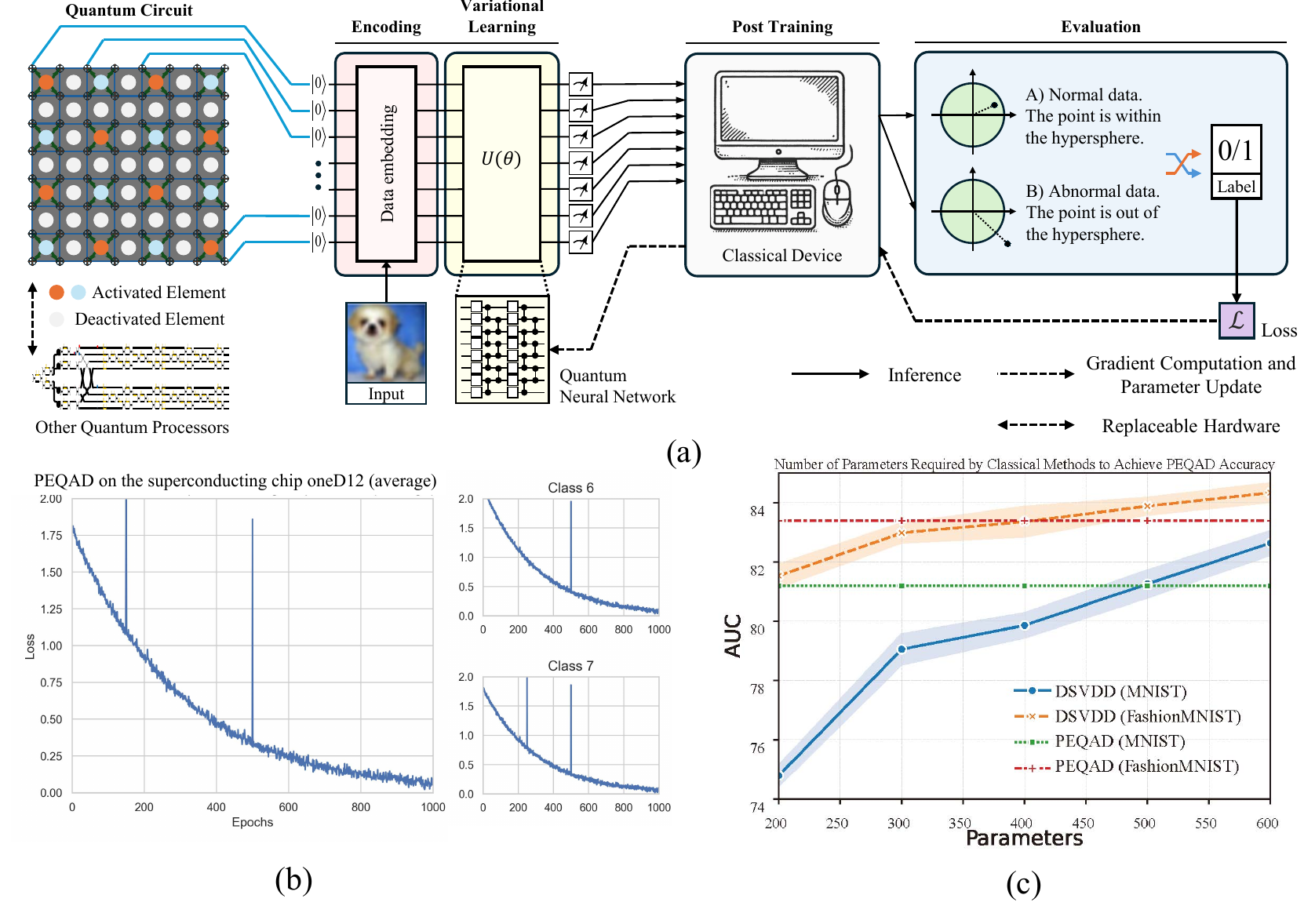}%
\caption{ (a) Conceptual diagram of deploying PEQAD on a quantum processor. (b) The optimization process of PEQAD on the superconducting processor oneD12 shows some fluctuations due to hardware instability. Despite these, the final loss is minimized effectively. (c) The parameter efficiency comparison between PEQAD on the superconducting quantum processor and DSVDD on a conventional computer. Dashed lines represent PEQAD hardware results using 16 parameters on MNIST or FashionMNIST, while solid lines show DSVDD performance as parameters increase. DSVDD requires significantly more parameters to match PEQAD performance.}
\label{fig: PEQAD on quantum processor}
\end{figure*}
The implementation of PEQAD on the quantum hardware platform is shown in Fig. \ref{fig: PEQAD on quantum processor} (a). To perform the VQC, each qubit is first initialized in the ground state, and encoding is applied in the pink region of the quantum processor to encode the input data. In the yellow region, the circuit is parameterized with single-qubit rotations and entangling gates between neighbouring qubits to create a trainable variational quantum circuit. After executing the VQC, we apply the pre-processing step of PEQAD by calculating the anomaly detection score based on the output quantum states. The results are then compared to a pre-determined threshold to classify the states, completing the implementation of PEQAD on the quantum processor.

We compare the performance of PEQAD across different quantum platforms, including emulators and the real superconducting quantum processor, namely the VQC emulator, the superconducting emulator, and the actual superconducting quantum processor oneD12. The superconducting emulator is implemented to replicate its respective chip structure accurately. In contrast, the VQC emulator is designed to resemble the general configuration of the hardware setups closely.
For the VQC emulation, we employed two different structures: one using a global $4$-qubit unitary with amplitude embedding (denoted as VQC) and one using rotation embedding (denoted as VQC (RE)). VQC and VQC (RE) are designed to mirror the configurations of the superconducting processor oneD12 closely but using different encoding methods. This approach allows us to evaluate the algorithm's adaptability to different quantum hardware structures and ansätze. 

All experiments were conducted under consistent conditions to ensure fairness, each using $100$ randomly selected normal samples for training and another $100$ for testing. All emulators were allocated computational resources equivalent to $4$ qubits. The MNIST and FashionMNIST datasets, originally consisting of 196 dimensions (after normalization and first initialization), were downsampled to 16 using a pooling technique. This reduction was done to facilitate efficient data encoding on the quantum hardware, which is constrained by the number of available qubits and their connectivity.  The training was performed over $200-1000$ epochs with a learning rate of $1\times10^{-2}$. For the optimization of PEQAD on the quantum processor, we utilized Adam and simultaneous perturbation stochastic approximation \cite{spall1992multivariate}. We continue to use AUC (Area Under the Curve) as the evaluation metric in this part.

\begin{table}[!htbp]
    \centering
    \caption{Results on the superconducting processor and the corresponding emulators. }
    \resizebox{1\linewidth}{!}{
    \begin{tabular}{cccccc}
    \toprule
     Dataset & Device & $Q$ & $P$ & AUC  \\
    \midrule
     \multirow{7}{*}{MNIST}      
     &VQC&	4&16	&84.8\% \\
     &oneD12 emulator&	4&16	&83.2\% \\
    &oneD12 processor&	4&16	&81.2\% \\
    &VQC (RE)&	4&16	&75.6\% \\
    &oneD12 processor (RE)&	4&16	&73.4\% \\

    &DSVDD &/ &500+&81.2\%\\
         &DSVDD &/ &600+&83.6\%\\

     \midrule
      \multirow{7}{*}{FashionMNIST}	
     &VQC &	4&16	&87.5\% \\

     &oneD12 emulator&	4&16	&85.2\% \\
     &oneD12 processor&	4&16	&83.4\% \\
     &VQC (RE)&	4&16	&82.1\% \\
    &oneD12 processor (RE)&	4&16	&77.4\% \\

         &DSVDD &/ &300+&83.4\%\\
             &DSVDD &/ &800+&85.3\%\\

    \bottomrule
    \end{tabular}}
\label{table: hardware results}
\end{table}

The experimental and emulation results of PEQAD are summarized in Table \ref{table: hardware results}. We observed that the performance of PEQAD on the superconducting processor is generally comparable to that obtained using emulators, including the superconducting emulator and VQC emulator.
On the MNIST dataset, PEQAD achieved an accuracy of $81.2\%$ on the superconducting processor, closely matching the $84.8\%$ accuracy obtained with the VQC emulator and surpassing the $75.6\%$ achieved with VQC (RE). On the FashionMNIST dataset, the accuracy reached $83.4\%$ on the superconducting processor, aligning well with the $87.5\%$ from VQC and $82.1\%$ from VQC (RE). Moreover, rotation encoding has been observed to markedly diminish the accuracy of recognition, both on the emulator and on the hardware. This outcome is likely attributable to the markedly insufficient exploitation of Hilbert space by this encoding method. Overall, despite a slight decline in accuracy when implemented on actual quantum hardware compared to emulators, the results from the superconducting processor demonstrated favourable performance, highlighting PEQAD's robustness on the real quantum platform. The observed discrepancies between hardware and emulation results can likely be attributed to the noisy conditions typical of NISQ devices and the limited number of qubits, which constrains the quantum model's expressivity. Importantly, PEQAD performed consistently well without any error mitigation techniques, demonstrating its resilience in realistic quantum computing environments.

It is important to note that our experiments indicate that PEQAD can achieve competitive performance with a much smaller number of parameters compared to its classical counterpart, DSVDD. 
To provide a fair comparison between PEQAD and its conventional counterpart, DSVDD, experiments and emulations were conducted under identical conditions, differing only in the number of model parameters. As illustrated in Fig. \ref{fig: PEQAD on quantum processor} (c), we progressively augmented the number of parameters in DSVDD, commencing from a modest value until its performance reached a level commensurate with PEQAD. DSVDD necessitates a parameter count that is more than a few tens of times greater than required to achieve the results of our 16-parameter experiment in hardware. The specific parameter numbers are presented in the accompanying table.  As shown in Table \ref{table: hardware results}, the findings demonstrated that DSVDD necessitated a minimum of 300 parameters to attain comparable recognition accuracy, whereas PEQAD only required four qubits and 16 learnable parameters. 

This notable reduction in the number of required parameters exemplifies the parameter efficiency of PEQAD, even within the constraints of current quantum hardware, such as limited gate depth and qubit count. A reduction in the number of parameters also results in a reduction in the complexity and difficulty of the training process, which is a crucial factor for machine learning algorithms. Moreover, the capacity to attain robust performance with a reduced number of parameters suggests that our approach can readily scale as quantum hardware advances, ultimately facilitating the utilisation of more sophisticated and profound quantum models. These insights highlight the potential of PEQAD in real-world quantum machine learning tasks and provide guidance for the design of future experiments aimed at further exploring the capabilities of quantum computing.

Furthermore, during the experiments on the superconducting processor oneD12, as shown in Fig. \ref{fig: PEQAD on quantum processor} (b), we observed some fluctuations in the loss function during optimization, which may be due to hardware instability inherent in the superconducting quantum device. Nevertheless, these fluctuations did not affect the final optimization result, largely due to the algorithm's incorporation of post-processing techniques such as regularization, which conferred robustness against such perturbations.

All the findings above underscore the practical potential of PEQAD and suggest that QML methods have the potential to be effectively implemented on quantum hardware. The decision to forgo error mitigation is deliberate, as we aimed to highlight the inherent noise tolerance of QML algorithms. This demonstrates the progress made in quantum machine learning and indicates that QML methods, when carefully designed, may be well-suited for the NISQ era, where noise and limited quantum circuit depth remain significant challenges.

\section{\label{sec:conclusion}Conclusion}
In this work, we present several contributions that have advanced the field of quantum machine learning for anomaly detection. First, we develop PEQAD, a novel quantum machine learning algorithm designed to be compatible with the NISQ device while effectively handling general image datasets. To establish a theoretical foundation, we provide a comprehensive examination of its expressivity, which offers a basis for its implementation on quantum hardware. We further validate the effectiveness and robustness of PEQAD by evaluating it on practical datasets, including different benchmarks, where it shows superior performance compared to various classical methods.  Furthermore, we successfully implemented PEQAD on the superconducting quantum device, demonstrating the feasibility of quantum-based anomaly detection and highlighting the potential benefits of reduced parameter requirements in quantum computing for this domain. By bridging quantum machine learning and anomaly detection, we demonstrate the use of quantum machine learning on widely relevant datasets, showcasing its diverse applications and reinforcing its role as a viable tool for addressing more real-world challenges. In summary, our method advances quantum anomaly detection by integrating parameter efficiency, scalability, and adaptability to general image datasets. Distinct from prior quantum anomaly methods, our approach demonstrates versatility and practicality across general image datasets, supported by theoretical analysis and validated through hardware experiments, emphasizing its parameter efficiency.

In the next phase of our work, we will focus on developing robust statistical techniques for anomaly detection and deriving generalization bounds to enhance algorithm robustness and applicability. We plan to incorporate error mitigation techniques and scale our experiments to larger and different quantum chips. We hope these improvements yield even more promising results, further showcasing the capabilities of PEQAD.  QML offers potential advantages in terms of accuracy, parameter efficiency, and scalability concerning training set size, all of which are particularly relevant advantages when applied to real-world classical datasets. 


\addvspace{1cm}
\section{Acknowledge}
The authors acknowledge funding support from the UK Engineering and Physical Sciences Research Council (EPSRC) under the project ``Software Environment for Actionable \& VVUQ-evaluated Exascale Applications (SEAVEA)'' (Grant No. EP/W007711/1).
 M.W. is grateful for the research studentship funded by the China Scholarship Council-UCL Joint Research Scholarship.   
The authors acknowledge Tim Weaving for his insightful comments and suggestions regarding the submission process. The authors also thank the members of the Qchemistry (Quantum Chemistry) Group in the Centre for Computational Science of University College London for their helpful input during our research. The authors acknowledge hardware support from the Institute of Quantum Information and Quantum Technology Innovation, Chinese Academy of Sciences in Shanghai, China. The authors also acknowledge using Pennylane AI and Qiskit software for this work.

\appendix




\section{Expressivity of Parameter-Efficient Quantum Anomaly Detection}\label{Appendix B}


The intuition of the proof is as follows. The proof of Theorem 1 employs the definition of the operator norm.
\begin{definition}[Operator norm]
Assume that $A$ is an $n \times n$ matrix. The operator norm of $A$ is defined as
\begin{equation*}
\|A\| = \sup_{\|x\|_2=1, x\in\mathbb{C}^n} \|A\bm{x}\|. 
\end{equation*}
Alternatively, $\|A\| = \sqrt{\lambda_1 (AA^{\dagger})}$, where $\lambda_i (AA^{\dagger})$ is the $i$-th largest eigenvalue of the matrix $AA^{\dagger}$.
\end{definition}

Besides the above definition, the proof of Theorem 1 leverages the following two lemmas. In particular, the first lemma allows us to employ the covering number of one metric space $ (\mathcal{H}_1, d_1)$ to bound the covering number of another metric space $ (\mathcal{H}_2, d_2)$.

\begin{lemma} [Lemma 5, \cite{barthel2018fundamental}]
\label{lem1}
Let $ (\mathcal{H}_1, d_1)$ and $ (\mathcal{H}_2, d_2)$ be metric spaces, and $f : \mathcal{H}_1 \rightarrow \mathcal{H}_2$ be $K$-Lipschitz continuous such that
\begin{equation*}
d_2 (f (\bm{x}), f (\bm{y}))\leq Kd_1 (\bm{x}, \bm{y}),\quad \forall \bm{x}, \bm{y} \in \mathcal{H}_1 .
\end{equation*}
Then, their covering numbers satisfy
\begin{equation*}
N (\mathcal{H}_2, \varepsilon, d_2)\leq N (\mathcal{H}_1, \varepsilon/K, d_1).
\end{equation*}
\end{lemma}

The second lemma presents the covering number of the operator group
\begin{equation*}
\mathcal{H}_{\mathrm{circ}} := \{ U (\theta)^\dagger O U (\theta) \,|\, \theta \in \Theta \},
\end{equation*}
where $U (\theta) = \prod_{i=1}^{k} u_i (\theta_i)$ and only $P$ parameters in $U (\theta)$ are trainable. The detail can be found in \cite{du2022efficient}.

\begin{lemma} [Lemma 2, \cite{du2022efficient}]
\label{lem2}
Following notations in Theorem 1,  suppose that the $q$-qubit \textit{ansätze} used contains a total of $q$ trainable parameters, with each gate acting on at most $m$ qubits. The $\varepsilon$-covering number for the operator group $\mathcal{H}_{\mathrm{circ}}$ with respect to the operator-norm distance satisfies:
\begin{equation*}
N (\mathcal{H}_{\mathrm{circ}}, \varepsilon, \|\cdot\|)\leq \left (\frac{7P\|O\|}{\varepsilon}\right)^{q^{2m}P}. 
\end{equation*}
\end{lemma}
where $P$ is the number of trainable parameters in the quantum circuit, $\|O\|$ is the operator norm of the observable $O$, and $q$ is the number of qubits. 

The covering number, $N(\mathcal{H}, \varepsilon, \|\cdot\|)$, provides a measure of the complexity of the hypothesis space $\mathcal{H}$. In the context of QML, this is particularly useful as it helps quantify the expressivity of quantum models by determining how well the hypothesis space can approximate different data distributions. The above formulas are derived following \cite{du2022efficient} and apply to all postprocessing-free VQAs, which means they can be used for theoretical calculations of expressivities for postprocessing-free variational quantum one-class algorithms such as VQOCC. 

Next, we proceed with the derivation to prove Theorem 1. Recall the definition of the hypothesis space $\mathcal{H}$ in Eq. (2) and Lemma 1. When $\mathcal{H}_1$ refers to the hypothesis space $\mathcal{H}$ and $\mathcal{H}_2$ refers to the unitary group $U(d^N)$, the upper bound of the covering number of $\mathcal{H}$, i.e., $N(\mathcal{H}, \varepsilon, d_2)$, can be derived by first quantifying the Lipschitz constant $K$ and then relating this to the covering number of the operator group $\mathcal{H}_{\mathrm{circ}}$ from Lemma 2. Under these observations, we now analyze the upper bound of the covering number $N(\mathcal{H}, \varepsilon, \|\cdot\|)$. Consider an arbitrary element $U$ from the function space $\mathcal{H}_{\mathrm{circ}}$. Since the function space $\mathcal{H}_{\mathrm{circ}}$ is bounded and can be covered by a finite number of elements concerning the operator norm $\|\cdot\|$, for any $\varepsilon > 0$, there exists an element $U_\varepsilon$ in the covering set such that the distance between $U$ and $U_\varepsilon$ satisfies
\begin{equation*}
\| U^\dagger O U - U_\varepsilon^\dagger O U_\varepsilon \| \leq \varepsilon.
\end{equation*}

This follows from the definition of the covering number, which ensures that for every element $U$ in $\mathcal{H}_{\mathrm{circ}}$, there exists an adjacent element $U_\varepsilon$ in the covering set whose distance from $U$ is bounded by $\varepsilon$. Therefore, we can derive the relation between $d_2 (S_{\mathrm{PEQAD}}(U), S_{\mathrm{PEQAD}} ( U_\varepsilon))$ and $d_1 (U (\theta)^\dagger OU (\theta), (U_\varepsilon (\theta))^\dagger O (U_\varepsilon (\theta)))$ as follows:
\begin{align*}
d_2(S_{\mathrm{PEQAD}}(U), S_{\mathrm{PEQAD}}(U_\varepsilon)) 
\\ =  \big| \nonumber 
(\mathrm{Tr}(U(\theta)^\dagger O U(\theta) \rho) - \boldsymbol{c})^2 - R^2 \nonumber \\
\quad \nonumber - \left( (\mathrm{Tr}(U_\varepsilon(\theta)^\dagger O U_\varepsilon(\theta) \rho) - \boldsymbol{c})^2 - R^2 \right) \big|
\end{align*}

where $\mathrm{Tr}(U(\theta)^\dagger OU(\theta) \rho)$ represents the expectation value of the observable $O$ concerning the quantum state $\rho$ transformed by the unitary $U(\theta)$. This expectation value is a critical component in determining the anomaly score $S_{\mathrm{PEQAD}}$, as it encodes the feature transformation performed by the VQC. $S_{\mathrm{PEQAD}}$ represents the anomaly score in the article. This is the final function output by PEQAD for determining whether the data is anomalous, which is the function we have learned. We abbreviate $\mathrm{Tr} (U (\theta)^\dagger OU (\theta) \rho)$ as $\mathrm{Tr}$ and $\mathrm{Tr} ( (U_\varepsilon (\theta))^\dagger O (U_\varepsilon (\theta)) \rho)$ as $\mathrm{Tr}\varepsilon$. Under this notation, the original equations can be written as follows:
\begin{align*}
d_2(S_{\mathrm{PEQAD}}(U), S_{\mathrm{PEQAD}}(U_\varepsilon)) 
\\ = \, \big| \nonumber
(\mathrm{Tr} - \boldsymbol{c})^2 - (\mathrm{Tr}_\varepsilon - \boldsymbol{c})^2 \nonumber \\
\quad \nonumber = (\mathrm{Tr} + \mathrm{Tr}_\varepsilon - 2\boldsymbol{c}) (\mathrm{Tr} - \mathrm{Tr}_\varepsilon) \\
\quad \nonumber \leq \| U(\theta)^\dagger O U(\theta) + U_\varepsilon(\theta)^\dagger O U_\varepsilon(\theta) - 2\boldsymbol{c} \| \\
\quad \nonumber \cdot \| U(\theta)^\dagger O U(\theta) - U_\varepsilon(\theta)^\dagger O U_\varepsilon(\theta) \| \cdot \mathrm{Tr}^2(\rho) \\
\quad \nonumber \leq |2\|O\| - 2\|\boldsymbol{c}\|| \cdot d_1(U(\theta)^\dagger O U(\theta), \\
\quad \nonumber U_\varepsilon(\theta)^\dagger O U_\varepsilon(\theta)),
\end{align*}
where the first inequality uses the triangle inequality once and the Cauchy-Schwartz inequality twice, and the second inequality employs the triangle inequality, $\mathrm{Tr}^2 (\rho) = 1$, $\|U\| = 1$ and also:
\begin{equation*}
\quad \| U^\dagger O U - U_\varepsilon^\dagger O U_\varepsilon \| = d_1(U^\dagger O U , U_\varepsilon^\dagger  O U_\varepsilon).
\end{equation*}

By applying the triangle inequality and the Cauchy-Schwarz inequality, we bound the difference between the anomaly scores $S_{\mathrm{PEQAD}}(U)$ and $S_{\mathrm{PEQAD}}(U_\varepsilon)$. The triangle inequality allows us to separate the difference into terms involving the operator norms of $U(\theta)^\dagger OU(\theta)$ and $U_\varepsilon(\theta)^\dagger O U_\varepsilon(\theta)$, which are further bounded using the properties of operator norms.
From here, we obtain the relationship between $ (\mathcal{H}, d_2)$ and $ (\mathcal{H}_{\mathrm{circ}}, d_1)$.

Next, we use Lemmas \ref{lem1} and \ref{lem2} and consider that
\begin{equation*}
d_2 (S_{\mathrm{PEQAD}} (\bm{x}), S_{\mathrm{PEQAD}} (\bm{y})) \leq Kd_1 (\bm{x}, \bm{y}), 
\end{equation*}
where $K=|(2\|O\|-2\|\boldsymbol{c}\|)|$.  In our case, $K$ is determined by the operator norm of $O$ and the $\bm{c}$. Specifically, we can derive that $K = |2\|O\| - 2\|\bm{c}\|| \leq 2\|O\| + 2\|\bm{c}\|$ by removing the absolute value. Now we have $K \leq 2\|O\| + 2\|\boldsymbol{c}\|$, and can further obtain:
\begin{equation*}
N (\mathcal{H}, \varepsilon, \|\cdot\|) \leq N (H_{\mathrm{circ}}, \varepsilon/K, \|\cdot\|) .
\end{equation*}

Substituting the value of $N (\mathcal{H}_{\mathrm{circ}}, \varepsilon, \|\cdot\|)$ from Lemma \ref{lem2}, we get
\begin{equation*}
N (\mathcal{H}, \varepsilon, \|\cdot\|) \leq \left (\frac{7P  (2\|O\| + 2\|\boldsymbol{c}\|)\|O\|}{\varepsilon}\right)^{q^{2m}P}.\
\end{equation*}

Therefore, we have obtained Theorem \ref{th}. 

For further consideration, VQC is a fully flexible component of our model. PEQAD can express the entire space that VQC is capable of. According to Lemma \ref{lem2}, assuming that the maximum theoretical covering number for VQC is:
\begin{equation*}
\sup N(\mathcal{H}_{\mathrm{circ}}, \varepsilon, \|\cdot\|) = \left( \frac{7P' \|O\|}{\varepsilon} \right)^{q^{2m}P'},
\end{equation*}
where \(P'\) represents the number of trainable parameters in the VQC model and \(q\) is the number of qubits.

Since PEQAD contains VQC as a subset, the supremum of the covering number of PEQAD must be greater than or equal to that of VQC. Therefore, we can have \(P'= P\) and establish the following inequality for the covering number of PEQAD:
\begin{equation*}
\sup N(\mathcal{H}, \varepsilon, \|\cdot\|) \geq \sup N(\mathcal{H}_{\mathrm{circ}}, \varepsilon, \|\cdot\|) = \left( \frac{7P \|O\|}{\varepsilon} \right)^{q^{2m}P}.
\end{equation*}

At the same time, the upper bound of the expressivity of PEQAD can be constrained by the following expression derived from the theoretical analysis of its function space:
\begin{equation*}
\sup N(\mathcal{H}, \varepsilon, \|\cdot\|) \leq \left( \frac{7P (2\|O\| + 2\|\boldsymbol{c}\|)\|O\|}{\varepsilon} \right)^{q^{2m}P}.
\end{equation*}

Therefore, the expressivity of PEQAD is bounded as follows:
\small
\begin{equation*}
\left( \frac{7P \|O\|}{\varepsilon} \right)^{q^{2m}P} \leq \sup N(\mathcal{H}, \varepsilon, \|\cdot\|) \leq \left( \frac{7P (2\|O\| + 2\|\boldsymbol{c}\|)\|O\|}{\varepsilon} \right)^{q^{2m}P}.
\end{equation*}

This is Theorem \ref{th2}. In particular, considering that $\|O\| = 1$ in our emulations and experiments, the upper bound of PEQAD’s expressivity is larger than that of its counterparts. Compared to post-processing-free variational quantum algorithms such as VQOCC, the expressivity of PEQAD benefits from the additional structure provided by the hypersphere constraint in the anomaly detection setting. Rather than reducing the hypothesis space, this constraint expands the expressive capacity of PEQAD, allowing it to capture a broader range of patterns. As a result, the expressivity of PEQAD is greater than other VQA-based anomaly detection methods, providing a more flexible framework for detecting complex anomalies. Intuitively, the expressivity of a quantum model, much like in classical machine learning, dictates its ability to model complex data distributions. While excessive expressivity can lead to overfitting in some cases, PEQAD's design balances this by utilizing a QNN architecture that enhances flexibility without succumbing to overfitting.

\section{ Expressivity of Parameter-Efficient Quantum Anomaly Detection on Noisy Devices}

In this section, we extend the expressivity analysis of PEQAD to quantum devices operating under noisy conditions. Similar to the proof of expressivity in the ideal case, we derive the covering number of PEQAD when the quantum operations are subject to noise.

We begin by recalling the covering number for PEQAD in the ideal (noise-free) case:
\begin{equation*}
    N(\mathcal{H}, \varepsilon, \|\cdot\|) \leq \left( \frac{7P(2\|O\| + 2\|\boldsymbol{c}\|)\|O\|}{\varepsilon} \right)^{q^{2m}P},
\end{equation*}

where \(\mathcal{H}\) denotes the hypothesis space spanned by the PEQAD model, \(P\) is the number of trainable parameters, and \(q\) is the number of qubits.

Next, we consider the noisy case. Under noisy conditions, the relation between the distance in the output states and the distance in the unitary operators becomes more complex, i.e., the depolarization channel \(\mathcal{E}_p(\cdot)\) is applied to each quantum gate in \(U(\cdot)\). Following the explicit form of the depolarization channel, the distance \(d_2\) between the noisy output states and the distance \(d_1\) between the unitary operators satisfy the following relation:
\begin{align*}
\nonumber d_2\Big( \mathrm{Tr}\big(O \mathcal{E}_p(U \rho U^\dagger)\big), \mathrm{Tr}\big(O \mathcal{E}_p(U_\varepsilon \rho U_\varepsilon^\dagger)\big) \Big)
\\ = \, \mathrm{Tr}\big( O \mathcal{E}_p(U \rho U^\dagger) - O \mathcal{E}_p(U_\varepsilon \rho U_\varepsilon^\dagger) \big) \nonumber \\
\quad \nonumber = (1 - p)^{N_g} \, \mathrm{Tr}\big( O ( U \rho U^\dagger - U_\varepsilon \rho U_\varepsilon^\dagger ) \big) \\
\quad \nonumber \leq (1 - p)^{N_g} \, \| U^\dagger O U - U_\varepsilon^\dagger O U_\varepsilon \| \, \mathrm{Tr}(\rho) \\
\quad \nonumber = (1 - p)^{N_g} \, \| U^\dagger O U - U_\varepsilon^\dagger O U_\varepsilon \| \\
\quad \nonumber = (1 - p)^{N_g} \, d_1(U, U_\varepsilon)
\end{align*}
where  \(N_g\) is the number of gates that are influenced by noise.

The second equality uses the depolarizing noise property. The inequality follows from applying the contractive property of the trace norm, and \(\mathrm{Tr}(\rho) = 1\) is used to simplify the expression.

This result indicates that the term \(K\) is:
\begin{equation*}
    K = (1 - p)^{N_g}.
\end{equation*}

Using Lemma \ref{lem2}, the covering number of the VQA under depolarizing noise is then upper bounded by:
\begin{equation*}
    N(\mathcal{H}_{\mathrm{circ}}, \varepsilon, \|\cdot\|) \leq (1 - p)^{N_g} \left( \frac{7 P \|O\|}{\varepsilon} \right)^{q^{2m} P}.
\end{equation*}
The bound reflects how the presence of depolarizing noise reduces the expressivity of the quantum variational circuit.

Following the same reasoning as in Theorem \ref{th}, we obtain that the expressivity of PEQAD in the presence of noise is bounded by:
\begin{equation*}
    N(\mathcal{H}, \varepsilon, \|\cdot\|) \leq (1 - p)^{N_g} \left( \frac{7P(2\|O\| + 2\|\boldsymbol{c}\|)\|O\|}{\varepsilon} \right)^{q^{2m}P},
\end{equation*}
where \(p\) represents the noise rate. This equation quantifies the effect of noise on the expressive power of PEQAD and offers insight into its limitations in practical, noisy quantum devices.

In this way, we derive Theorem \ref{th3}. This bound demonstrates that the expressivity of PEQAD is reduced in noisy environments due to the noise factor \((1 - p)^{N_g}\). As the noise level increases (i.e., as \(p\) approaches 1), the effective hypothesis space is reduced, limiting the model's ability to capture complex patterns.

\section{Expressivity  of the Classical Machine Learning Counterparts}\label{Classical model expressivity bound}

To compare the expressivity of PEQAD with classical counterparts, we provide the detailed derivation of the covering number bound for classical neural networks based on Bartlett et al.'s \cite{NIPS2017_b22b257a} Theorem 3.3. The covering number bound for a classical neural network is given by:
\begin{equation*}
\ln N (\mathcal{H}_c, \varepsilon, \|\cdot\|_2) \leq \frac{t^2 \ln(2W^2)}{\varepsilon^2} \left( \prod_{j=1}^{L} s_j^2 \rho_j^2 \right) \left( \sum_{i=1}^{L} \left( \frac{b_i}{s_i} \right)^{2/3} \right)^3,
\end{equation*}
where \( t \) is the norm of the data matrix, \(p\) represents the noise rate. Since the training set is fixed in our comparison emulations and experiments, we treat the norm of the data matrix as a constant \( t \) to simplify our analysis. \( s_i \) represents the spectral norm bound of the weight matrix at layer \( i \). \( b_i \) represents the (2,1) matrix norm bound of the weight matrix at layer \( i \). \( \rho_i \) represents the Lipschitz constant of the activation function at layer \( i \). \( W \) is the maximum dimension of the matrices in the network layers. We can express \( s_i \) and \( b_i \) regarding network structure to further analyze their impact.

Spectral norm bound \( s_i \) can be estimated as:
\begin{equation*}
s_i \leq \sqrt{m_i n_i} \, \|W_i\|_{\infty},
\end{equation*}
where \( m_i \) and \( n_i \) are the output and input dimensions of the weight matrix \( W_i \) at layer \( i \), respectively, and \( \|W_i\|_{\infty} \) represents the maximum absolute value of the elements in the weight matrix \( W_i \).

For the (2,1) norm bound \( b_i \) of the weight matrix at layer \( i \):
\begin{equation*}
b_i \leq n_i \sqrt{m_i} \, \|W_i\|_{\infty}.
\end{equation*}

Substituting the upper bounds of \( s_i \) and \( b_i \) into the original covering number bound, we have:
\begin{eqnarray*}
\ln N (\mathcal{H}_c, \varepsilon, \|\cdot\|_2) &\leq& \frac{t^2 \ln(2W^2)}{\varepsilon^2} \left( \prod_{j=1}^{L} \left( \sqrt{m_j n_j} \, \|W_j\|_{\infty} \right)^2 \rho_j^2 \right) \nonumber\\
& & \times \left( \sum_{i=1}^{L} \left( \frac{n_i \sqrt{m_i} \, \|W_i\|_{\infty}}{\sqrt{m_i n_i} \, \|W_i\|_{\infty}} \right)^{2/3} \right)^3.
\end{eqnarray*}

Further simplifying:
\begin{eqnarray*}
\ln N (\mathcal{H}_c, \varepsilon, \|\cdot\|_2) &\leq& \frac{t^2 \ln(2W^2)}{\varepsilon^2} \left( \prod_{j=1}^{L} m_j n_j \|W_j\|_{\infty}^2 \rho_j^2 \right) \nonumber\\
& & \times \left( \sum_{i=1}^{L} \left( \frac{\sqrt{m_i}}{\sqrt{n_i}} \right)^{2/3} \right)^3.
\end{eqnarray*}

Finally, we can represent \( W \) as the maximum of \( m_j \times n_j \):
\begin{equation*}
W = \max(m_j \times n_j).
\end{equation*}

We also note that the Lipschitz constant of the activation functions used in our networks is always less than or equal to 1. For example, the ReLU activation function has a Lipschitz constant of 1, commonly used in deep networks. Thus, the covering number bound becomes:
\begin{eqnarray*}
\ln N (\mathcal{H}_c, \varepsilon, \|\cdot\|_2) &\leq& \frac{t^2 \left( \ln(2) + 2 \ln(\max(m_j \times n_j)) \right)}{\varepsilon^2} \nonumber\\
& & \times \left( \prod_{j=1}^{L} m_j n_j \|W_j\|_{\infty}^2 \right) \left( \sum_{i=1}^{L} \left( \frac{\sqrt{m_i}}{\sqrt{n_i}} \right)^{2/3} \right)^3.
\end{eqnarray*}

Thus, we obtained Theorem \ref{th:classical}.

\section{Analysis and Comparison of PEQAD and Classical Methods}\label{Analysis and Comparison of PEQAD and classical methods}

In this section, we provide a detailed comparison of the expressivity bounds of PEQAD and classical deep machine learning anomaly detection models, focusing on their respective covering number bounds. Recall that the expressivity of PEQAD can be characterized by the following covering number bound:
\begin{equation*}
N (\mathcal{H}_q, \varepsilon, |\cdot|) \leq \left( \frac{7P (2\|O\| + 2\|\boldsymbol{c}\|)\|O\|}{\varepsilon} \right)^{q^{2m}P}.
\end{equation*}

Taking the natural logarithm of both sides, we obtain:
\begin{equation*}
\ln N (\mathcal{H}_q, \varepsilon, |\cdot|) \leq q^{2m} P \cdot \ln \left( \frac{7P (2\|O\| + 2\|\boldsymbol{c}\|)\|O\|}{\varepsilon} \right).
\end{equation*}

This logarithmic form facilitates a direct comparison with the expressivity bound of classical models. For a classical neural network consisting of two layers, where the first layer has an input dimension of 40 and an output dimension of 10, and the second layer has an input dimension of 10 and an output dimension of 2, the covering number bound can be expressed as:
\begin{align*}
\ln N (\mathcal{H}_c, \varepsilon, \|\cdot\|_2) \leq \frac{t^2 \left( \ln(2) + 2 \ln(400) \right)}{\varepsilon^2} \times 8000 \\\times \|W_1\|_{\infty}^2 \|W_2\|_{\infty}^2 \times \left( 2^{-2/3} + 5^{-1/3} \right)^3.
\end{align*}

This bound indicates that the expressivity of classical neural networks depends on a polynomial relationship between various network parameters, including input-output dimensions, the number of layers, and the number of trainable parameters.

For PEQAD with four qubits (\( q = 4 \)), 16 trainable parameters (\( P = 16 \)), and a maximum gate interaction involving two qubits (\( k = 2 \)), the covering number bound can be simplified to:
\begin{equation*}
\ln N (\mathcal{H}_q, \varepsilon, |\cdot|) \leq 4096 \cdot \ln \left( \frac{7 \times 16 (2\|O\| + 2\|\boldsymbol{c}\|)\|O\|}{\varepsilon} \right).
\end{equation*}

Although it is difficult to compare the two directly, this bound shows that the expressivity of PEQAD scales in a polynomial fashion, similar to the classical model. However, PEQAD achieves this level of expressivity with significantly fewer parameters compared to the classical model. In both models, expressivity is influenced by the number of parameters. Still, in different forms, the classical model's expressivity depends on input-output dimensions, layer size, and the magnitude of the weights, forming a complex polynomial relationship, while PEQAD's expressivity also has a polynomial dependence on the number of trainable parameters \( P \) and the number of qubits \( q \). Under the given conditions—4 qubits and 16 trainable parameters for PEQAD versus a classical two-layer network—the expressivity bounds can be comparable. This suggests that PEQAD can achieve an expressivity similar to classical networks, even with fewer parameters.
Moreover, the true strength of PEQAD lies in its potential to leverage quantum properties such as entanglement and superposition, which might provide practical advantages in certain scenarios, particularly for high-dimensional problems or complex data distributions. Thus, we illustrate the comparison of the expressivity of our algorithm and a classical deep machine learning anomaly detection through this specific example, which is part of the conclusion. In the experimental section, we further validate the capability of our algorithm through empirical results.

\section{Quantitative Comparison with Typical Parameters}
To provide a more tangible and straightforward comparison, as suggested by the reviewer, we now quantify the expressivity bounds derived in our theoretical analysis. We substitute typical parameter values—drawn from our hardware experiments and common deep learning architectures—to illustrate the claim of parameter efficiency. Our goal is to estimate the number of classical parameters required for a network to achieve a theoretical expressivity comparable to that of our small-scale quantum model.

We begin with the expressivity bound for PEQAD, as given in Theorem 1. We consider the model implemented on our quantum hardware, which is characterized by the following parameters:
Number of qubits, $q=4$.
Number of trainable parameters, $P=16$.
Maximum gate locality, $m=2$ (for our two-qubit CNOT gates).
Furthermore, we make the standard assumptions that the observable is a Pauli operator, thus its operator norm $\|O\| = 1$, and that the hypersphere's center norm $\|\boldsymbol c\|$ is on the order of 1, which is reasonable for normalized data. Substituting these values into the logarithmic covering number bound:
\begin{align*}
    \ln N(\mathcal{H}_{q},\epsilon,|\cdot|) &\le q^{2m}P\cdot \ln\left(\frac{7P(2\|O\|+2\|\boldsymbol c\|)\|O\|}{\epsilon}\right) \nonumber \\
    &\le 4^{2 \times 2} \times 16 \cdot \ln\left(\frac{7 \times 16(2 \cdot 1+2 \cdot 1)\cdot 1}{\epsilon}\right) \nonumber \\
    &\le 256 \times 16 \cdot \ln\left(\frac{448}{\epsilon}\right) \nonumber \\
    &\le 4096 \cdot \ln\left(\frac{448}{\epsilon}\right).
\end{align*}
This result indicates that the theoretical expressivity of our 16-parameter quantum model scales with a pre-factor of approximately 4096.

Next, we analyze the bound for a classical deep network (DSVDD) from Theorem 4. Following the reviewer's suggestion, we consider a typical 5-layer network ($L=5$). We define an architecture where the input dimensionality matches our quantum encoding ($n_1=16$) and the network width is inspired by the suggestion of $W=256$:
\begin{center}
    $16 \xrightarrow{n_1=16, m_1=256} 256 \xrightarrow{n_2=256, m_2=128} 128 \xrightarrow{n_3=128, m_3=64} 64 \xrightarrow{n_4=64, m_4=32} 32 \xrightarrow{n_5=32, m_5=10} 10$.
\end{center}
The total number of parameters for this network exceeds 45,000. We again assume normalized data and weights, such that the data norm $t \approx 1$ and the weight matrix norm $\|W_{j}\|_{\infty} \approx 1$. The classical expressivity bound is highly sensitive to the term $\prod_{j=1}^{L}m_{j}n_{j}$:
\begin{align*}
    \prod_{j=1}^{L}m_{j}n_{j} &= (256 \times 16) \times (128 \times 256) \times (64 \times 128) \nonumber \\
    &\quad \times (32 \times 64) \times (10 \times 32) \nonumber \\
    &\approx (4.1 \times 10^{3}) \times (3.3 \times 10^{4}) \times (8.2 \times 10^{3}) \nonumber \\
    &\quad \times (2.0 \times 10^{3}) \times (3.2 \times 10^{2}) \nonumber \\
    &\approx 7.1 \times 10^{17}.
\end{align*}
Even without calculating the other pre-factors, it is evident that this product term, which is directly related to the network's parameter count and dimensions, results in an astronomically large value for the expressivity bound.

This illustrative calculation reveals a stark contrast in how the models' parameters contribute to their theoretical expressivity. To achieve an expressivity bound ($\ln N_c$) on the same order of magnitude as that of the quantum model ($\ln N_q \sim O(10^3)$), the classical network relies on a massive parameter count, which is reflected in the extremely large product-of-dimensions term. This provides a quantitative, albeit approximate, justification for our claim of parameter efficiency. It demonstrates how a quantum model, by leveraging the high dimensionality of Hilbert space (captured by the $q^{2mP}$ term), may achieve a rich expressive capacity with substantially fewer parameters than a classical deep network. This theoretical insight aligns well with our empirical findings, where the 16-parameter PEQAD performed comparably to a classical DSVDD requiring over 300 parameters.

\section{Barren Plateaux Phenomenon for PEQAD}

In this section, we will show that even if our algorithm increases the expressive power using post-processing, it does not accelerate the barren plateaux (BP) phenomenon \cite{mcclean2018barren}.  We provide the detailed derivation showing that the PEQAD loss function \((E(\theta) - \boldsymbol{c})^2 - R^2\) exhibits the BP phenomenon under a 1-design unitary distribution. While the expected gradient vanishes, the presence of the factor \((E(\theta) - \boldsymbol{c})\) can modulate the rate of decay, which provides a nuanced understanding of PEQAD's expressivity without exacerbating the BP problem. This appendix further provides the theoretical derivation supporting the claim in Section 5 that the PEQAD ansatz does not exhibit BP on our physical circuit.

The PEQAD loss function is defined as:
\[
L(\theta) = (E(\theta) - \boldsymbol{c})^2 - R^2,
\]
where \(R^2\) is a constant. Since \(R^2\) does not depend on \(\theta\), it does not contribute to the gradient. Thus, for gradient analysis, we simplify the loss function to:
\[
L(\theta) = (E(\theta) - \boldsymbol{c})^2.
\]

Here, \(E(\theta)\) is the expectation value:
\[
E(\theta) = \langle 0| U^\dagger(\theta) O U(\theta) |0\rangle.
\]

The gradient of \(L(\theta)\), with respect to the \(k\)-th parameterized quantum gate, is given by:
\[
\partial_k L(\theta) = 2(E(\theta) - \boldsymbol{c}) \cdot \partial_k E(\theta).
\]

Here, the gradient of \(E(\theta)\), with respect to the \(k\)-th parameterized quantum gate, is expressed as:
\[
\partial_k E(\theta) = \frac{\partial}{\partial \theta_k} \langle 0| U^\dagger(\theta) O U(\theta) |0\rangle.
\]

Using the parameterized unitary decomposition \(U(\theta) = U_+ U_-\), where \(U_+\) represents the product of gates from the \((k+1)\)-th to the last parameterized gate, and \(U_-\) represents the product of gates from the first to the \(k\)-th parameterized gate, we have:
\[
V_k = \frac{\partial U_+}{\partial \theta_k} U_+^\dagger.
\]

The derivative of \(E(\theta)\) is given by:
\[
\partial_k E(\theta) = \frac{\partial}{\partial \theta_k} \langle 0| U^\dagger(\theta) O U(\theta) |0\rangle.
\]

Using the decomposition \(U(\theta) = U_+ U_-\), we write:
\[
\frac{\partial U(\theta)}{\partial \theta_k} = U_+ \frac{\partial U_-}{\partial \theta_k} + \frac{\partial U_+}{\partial \theta_k} U_-.
\]

Substituting this into \(\partial_k E(\theta)\), we have:
\[
\partial_k E(\theta) = \langle 0| \frac{\partial U^\dagger(\theta)}{\partial \theta_k} O U(\theta) |0\rangle + \langle 0| U^\dagger(\theta) O \frac{\partial U(\theta)}{\partial \theta_k} |0\rangle.
\]

For the first term:

\[
\frac{\partial U^\dagger(\theta)}{\partial \theta_k} = \frac{\partial U_-^\dagger}{\partial \theta_k} U_+^\dagger + U_-^\dagger \frac{\partial U_+^\dagger}{\partial \theta_k}.
\]

Thus:
\begin{align}
\nonumber \langle 0| \frac{\partial U^\dagger(\theta)}{\partial \theta_k} O U(\theta) |0\rangle 
=\, & \langle 0| \frac{\partial U_-^\dagger}{\partial \theta_k} U_+^\dagger O U_+ U_- |0\rangle \nonumber \\
& \nonumber + \langle 0| U_-^\dagger \frac{\partial U_+^\dagger}{\partial \theta_k} O U_+ U_- |0\rangle
\end{align}

For the second term:
\begin{align*}
\langle 0| U^\dagger(\theta) O \frac{\partial U(\theta)}{\partial \theta_k} |0\rangle 
=\, & \langle 0| U_-^\dagger U_+^\dagger O U_+ \frac{\partial U_-}{\partial \theta_k} |0\rangle \\
& + \langle 0| U_-^\dagger U_+^\dagger O \frac{\partial U_+}{\partial \theta_k} U_- |0\rangle
\end{align*}

Combining these terms and simplifying using the commutator, we obtain:
\[
\partial_k E(\theta) = i \langle 0| U_-^\dagger [V_k, U_+^\dagger O U_+] U_- |0\rangle,
\]
where \(V_k = \frac{\partial U_+}{\partial \theta_k} U_+^\dagger\).

Recall the loss function \(L(\theta) = (E(\theta)-c)^2\), where:
\[
\partial_k L(\theta) = 2(E(\theta)-c)\partial_k E(\theta).
\]

Consider the parameterized unitary \(U(\theta) = U_+ U_-\), with the distribution defined by \(p(U_+)\) and \(p(U_-)\). For the overall unitary distribution \(p(U)\), we write:
\[
p(U) = \int dU_+ p(U_+) \int dU_- p(U_-) \delta(U_+U_- - U),
\]
where \(\delta\) is the delta function over the unitary group, ensuring \(U = U_+ U_-\).

The expectation of \(\partial_k L(\theta)\) over the unitary group distribution is:
\[
\langle \partial_k L(\theta) \rangle = \int dU p(U) \partial_k L(\theta).
\]
Substituting the decomposition of \(p(U)\):
\begin{align*}
\langle \partial_k L(\theta)\rangle 
= \int dU_+\, p(U_+) \int dU_-\, p(U_-) \int dU\, \delta(U_+ U_- - U) \\
\cdot\, 2(E(\theta) - c)\, \partial_k E(\theta)
\end{align*}
Since \(\delta(U_+ U_- - U)\) ensures \(U = U_+ U_-\), this simplifies to:
\[
\langle \partial_k L(\theta)\rangle = \int dU_+ p(U_+) \int dU_- p(U_-) 2(E(\theta)-c) \partial_k E(\theta).
\]

Substituting the expression for \(\partial_k E(\theta)\):
\begin{align*}
\langle \partial_k L(\theta) \rangle 
= \int dU_+\, p(U_+) \int dU_-\, p(U_-) \cdot\, 2(E(\theta) - c)\, \cdot\, i \\
\cdot\, \langle 0| U_-^\dagger [V_k, U_+^\dagger O U_+] U_- |0\rangle
\end{align*}

With \(\rho_- = U_- |0\rangle\langle0|U_-^\dagger\), we rewrite:
\begin{align*}
\langle \partial_k L(\theta) \rangle 
=\, 2i \int dU_+\, p(U_+) \int dU_-\, p(U_-) \cdot\, (E(\theta) - c) \\
\cdot\, \operatorname{Tr}\left( [V_k, U_+^\dagger O U_+]\, \rho_- \right)
\end{align*}
Recall:
\[
E(\theta) = \operatorname{Tr}(O U_+ \rho_- U_+^\dagger).
\]

Case 1: if \(U_-\) is a 1- design unitary matrix: Averaging over \(U_-\), we use the 1-design property:
\[
\int dU_- p(U_-) \rho_- = \frac{I}{D}, 
\] 
where $D = 2^q$  is the Hilbert space dimension. The averaged sense, \(\rho_- \to I/D\). Substituting, we have:
\begin{align*}
\langle \partial_k L(\theta) \rangle 
=\, 2i \int dU_+\, p(U_+) \cdot \left( \frac{\operatorname{Tr}(O)}{D} - c \right) \\
\cdot\, \operatorname{Tr}\left( [V_k, U_+^\dagger O U_+] \cdot \frac{I}{D} \right)
\end{align*}

Since the commutator's trace vanishes:
\[
\operatorname{Tr}([V_k, X]) = 0, \quad \forall X.
\]

Thus:
\[
\langle \partial_k L(\theta) \rangle = 0.
\]

Case 2: \(U_+\) is a 1-design unitary matrix. Using the 1-design property for \(U_+\):
\[
\int dU_+ p(U_+) U_+^\dagger O U_+ = \frac{\operatorname{Tr}(O)}{D}I.
\]

Substituting, we have:
\begin{align*}
\langle \partial_k L(\theta) \rangle 
=\, 2i \int dU_-\, p(U_-) \cdot \left( \frac{\operatorname{Tr}(O)}{D} - c \right) \\
\cdot\, \operatorname{Tr}\left( [V_k, U_+^\dagger O U_+] \cdot \frac{\operatorname{Tr}(O)}{D} I \right)
\end{align*}

The commutator \([V_k, U_+^\dagger O U_+]\) vanishes, as \([V_k, I] = 0\). 

Thus:
\[
\langle \partial_k L(\theta) \rangle = 0.
\]

Combining these results, we have: As long as any portion of the circuit approaches a 1-design, the gradient expectation will vanish, resulting in \(\langle \partial_k L(\theta) \rangle = 0\).
Although the expected gradient vanishes, the presence of \((E(\theta) - \boldsymbol{c})\) modulates the gradient's decay rate. One of the key observations is that when \(|E(\theta) - \boldsymbol{c}| > 1\), the decay of the gradient is slower, which helps maintain a balance between expressivity and optimization. In our emulations, this condition is satisfied in most cases, indicating that PEQAD generally operates in a regime where the gradient decay is mitigated, allowing for enhanced expressivity without significantly exacerbating the BP phenomenon.
Even in scenarios where \(|E(\theta) - \boldsymbol{c}| \leq 1\), our results indicate that the algorithm does not exhibit a substantial acceleration of the BP issue. 

Below, we provide a mathematical derivation to extend the scaling results of the variance of \(\partial_k E(\theta)\) under 2-design Haar distributions to the case of \(\partial_k L(\theta)\), where 
\[
L(\theta) = (E(\theta) - \boldsymbol{c})^2.
\]

The variance of \(\partial_k L(\theta)\) is:
\[
\operatorname{Var}[\partial_k L] = \langle (\partial_k L)^2 \rangle - \langle \partial_k L \rangle^2.
\]

For the first term, 
\[
(\partial_k L)^2 = 4(E(\theta) - \boldsymbol{c})^2 \cdot (\partial_k E(\theta))^2.
\]

Taking the expectation over the 2-design distributions for \(U_-\) and \(U_+\), we write:
\[
\langle (\partial_k L)^2 \rangle = 4\langle (E(\theta) - \boldsymbol{c})^2 \cdot (\partial_k E(\theta))^2 \rangle.
\]

From standard results for 2-design distributions:
The variance of \(\partial_k E(\theta)\) scales as \cite{mcclean2018barren}:
\[
\operatorname{Var}[\partial_k E(\theta)] = O\left(\frac{1}{D^2}\right).
\]

For \((E(\theta) - \boldsymbol{c})^2\), the second moment remains constant so that it can be bounded as:
\[
\langle (E(\theta) - \boldsymbol{c})^2 \rangle = O\left(1\right).
\]


Substituting:
\[
\langle (E(\theta) - \boldsymbol{c})^2 \rangle = O\left(1\right), \quad \langle (\partial_k E(\theta))^2 \rangle = O\left(\frac{1}{D^2}\right).
\]

Thus, we find the worst case is that:
\[
\langle (\partial_k L)^2 \rangle = O\left(1 \cdot \frac{1}{D^2}\right) = O\left(\frac{1}{D^2}\right).
\]

For the second term:
\[
\langle \partial_k L \rangle = \langle 2(E(\theta) - \boldsymbol{c}) \cdot \partial_k E(\theta) \rangle.
\]

From the properties of 2-design unitaries:
\[
\langle \partial_k E(\theta) \rangle = 0.
\]

Since \((E(\theta) - \boldsymbol{c})\) is independent of the specific gradient distribution \(\partial_k E(\theta)\):
\[
\langle (E(\theta) - \boldsymbol{c}) \cdot \partial_k E(\theta) \rangle = \langle (E(\theta) - \boldsymbol{c}) \rangle \cdot \langle \partial_k E(\theta) \rangle = 0.
\]

Thus:
\[
\langle \partial_k L \rangle = 0, \quad \langle \partial_k L \rangle^2 = 0.
\]

The variance of \(\partial_k L(\theta)\) scales as:
\[
\operatorname{Var}[\partial_k L] = O\left(\frac{1}{D^2}\right).
\]

Since \(D = 2^q\), the scaling becomes:
\[
O\left(\frac{1}{D^2}\right) = O\left(\frac{1}{2^{2q}}\right).
\]

Using Chebyshev's inequality, we can bound the probability of observing significant deviations of \(\partial_k C\) from its mean:
\[
P\left(|\partial_k C - \langle \partial_k C \rangle| \geq c'\right) \leq \frac{\operatorname{Var}[\partial_k C]}{c'^2}.
\]

Given that:
\[
\operatorname{Var}[\partial_k C] = O\left(\frac{1}{2^{2q}}\right),
\]
we have:
\[
P\left(|\partial_k C| \geq c'\right) \leq \frac{O\left(\frac{1}{2^{2q}}\right)}{c'^2}.
\]

This result demonstrates that the gradient variance decays exponentially with the number of qubits \(q\), consistent with the barren plateaux phenomenon in normal variational quantum circuits. 

In conclusion, the PEQAD loss function retains the BP phenomenon inherent in its underlying variational structure under $t$-design ($t \geq 2$) unitary distributions. However, our earlier supplemental analysis demonstrated that the post-processing mechanism in PEQAD may increase the expressive capacity of the model, providing a larger hypothesis space for learning.  Crucially, as shown in this section, the enhanced expressive capacity of PEQAD does not significantly exacerbate the BP phenomenon. The loss function structure ensures manageable gradient decay, balancing expressivity and optimization stability. 
\begin{figure*}[!htbp]
    \centering
    \subfloat[]{
        \includegraphics[width=0.47\textwidth]{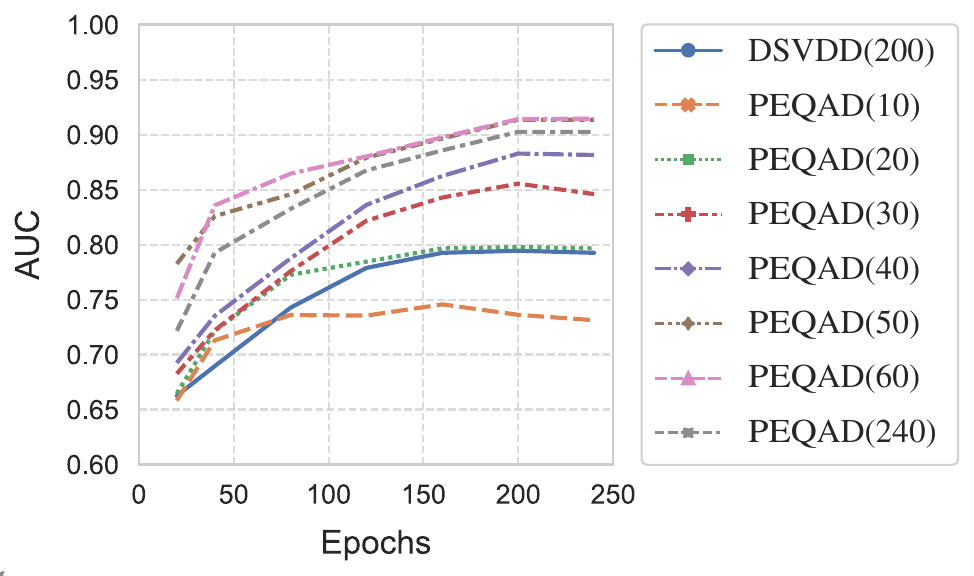}
        \label{fig:epoch}
    }
    \hfill
    \subfloat[]{
        \includegraphics[width=0.49\textwidth]{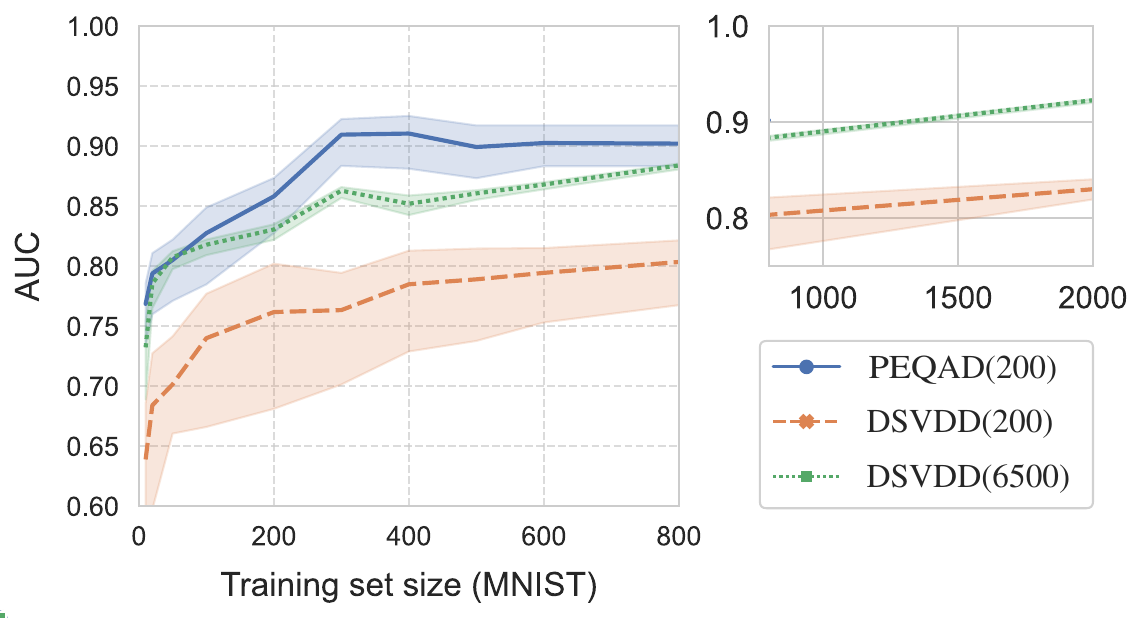}
        \label{fig:trainset}
    }
    \caption{Comparison of PEQAD and DSVDD on FashionMNIST and MNIST datasets. (a) shows the AUC trends with different trainable parameters on FashionMNIST. As the number of trainable parameters in PEQAD increases from 10 to 70, its AUC shows an upward trend. However, when the number of parameters becomes excessively large, leading to over-expressivity, optimisation becomes more challenging. Specifically, this is reflected in the AUC dropping at the same epoch. For instance, as shown in the figure, when the parameter count reaches 240, the AUC of PEQAD performs worse than with 60 parameters. (b) highlights the performance under varying training set sizes on MNIST. When the number of samples in the training set is small (less than 400 in this case), PEQAD outperforms DSVDD significantly, even though DSVDD has more trainable parameters.
    }
    \label{fig:combined}
\end{figure*}

The above analysis assumes Haar-random circuits drawn from a $t$-design ($t \geq 2$) distribution. This assumption is made to support generality, since different implementations of PEQAD may involve distinct variational ansätze. However, in this section, we also provide a refined derivation of the gradient behavior specific to our physical circuit structure (i.e., RY+CNOT brickwall ansatz), leveraging direct moment computation to show that barren plateaus do not occur under our design \cite{yao2025direct}. Our ansatz consists of alternating layers of fixed CNOT gates arranged in a brick-wall pattern and parameterized single-qubit $\mathrm{RY}(\theta)$ rotations, with one $\mathrm{RY}$ gate per qubit per layer. Let $q = 4$ be the number of qubits and $P = 16$ the total number of trainable parameters, where each layer contributes $q$ parameters. The measurement observable is a local operator $O = Z_j$ acting on qubit $j$, and the loss function is again defined as $L(\theta) = (E(\theta) - \boldsymbol{c})^2$, with $E(\theta) = \langle 0| U^\dagger(\theta) O U(\theta) |0\rangle$. The gradient of the loss with respect to any parameter $\theta_k$ is given by
\[
\partial_k L = 2(E(\theta) - \boldsymbol{c}) \cdot \partial_k E(\theta).
\]
To compute $\partial_k E(\theta)$, we consider the Heisenberg-evolved observable $U^\dagger(\theta) O U(\theta)$ and how it transforms under conjugation by the single-qubit gate $U_k(\theta_k) = \mathrm{RY}(\theta_k)$. For any Hermitian operator $A$ acting on $q$ qubits, if $U_k$ acts only on qubit $k$, then the expected conjugation over the rotation axis (assuming uniformly sampled RX, RY, RZ) yields:
\[
\mathbb{E}\left[ U_k^\dagger A U_k \right] = \frac{1}{3} \left( A + \mathbb{I}_k \otimes \operatorname{Tr}_k A \right),
\]
where $\operatorname{Tr}_k A$ denotes the partial trace over qubit $k$, and $\mathbb{I}_k$ is the identity operator acting on that qubit. Applying this recursively across all $q$ qubits, and noting the locality of the observable $O$, the expected gradient becomes \cite{yao2025direct}:
\[
|\langle \partial_k L \rangle| \propto \frac{1}{3^q}.
\]
This demonstrates that the expected gradient decays exponentially with the number of qubits $q$, but does not vanish entirely, in contrast to results derived under the Weingarten calculus assumption of full Haar randomness. Moreover, the variance of the gradient is shown to scale as
\[
\operatorname{Var}[\partial_k L] \propto \frac{p}{8^q \cdot P},
\]
where $p$ denotes the number of effective parameters that influence the measurement observable $O$, which remains small due to the locality of light-cone propagation through the layered CNOT structure. In our setting with $q = 4$ and $P = 16$, the expected gradient satisfies $|\langle \partial_k L \rangle| \sim 1.2 \times 10^{-2}$, while the variance is approximately $\sim 3.1 \times 10^{-6}$, since $p = O(1)$. These results confirm that the barren plateau phenomenon does not occur in our ansatz, as the gradients remain non-zero and statistically significant even in realistic quantum circuits, enabling effective optimization of the PEQAD objective.

This provides a viable path for improving variational quantum circuits under practical constraints, showcasing the robustness and adaptability of PEQAD in both theoretical and experimental contexts. 

\section{Analysis of the Theoretical Expressivity of PEQAD and Emulation Agreement}\label{Analysis and Comparison}


The analysis of the theoretical expressivity of PEQAD and its agreement with emulation results reveals several important insights. Based on the data presented in Figure \ref{fig:combined} and the theoretical analysis, we can observe that when the expressivity of PEQAD is relatively weak, i.e., when the target concept is excluded, training remains stable but yields unsatisfactory results, i.e., underfitting. In such cases, increasing the number of trainable parameters increases the expressivity. This leads to improved performance, as shown in the figure when we incrementally increased the number of trainable parameters from 10 to 70, resulting in an increase in AUC. Conversely, too much expressivity leads to a worse AUC, i.e., overfitting. This can be seen in the figure, where the PEQAD model with 60 trainable parameters outperforms the model with 240 parameters.
In summary, finding the right balance of expressivity in PEQAD is essential.  The results of our emulations underscore the importance of carefully tuning expressivity to achieve optimal performance in anomaly detection tasks. 

Furthermore, the results presented in Fig.~\ref{fig:epoch} and Fig.~\ref{fig:trainset} offer significant insights into the expressivity of QML compared to deep classical machine learning, as theoretically derived earlier. Classical neural networks are characterised by their strong expressivity, unlike conventional QML methods, which frequently exhibit a more constrained expressivity range during target function learning. This restricted range enhances trainability but concomitantly increases the probability of failing to fully capture the learning target. However, when the optimal solution is within reach, lower expressivity frequently translates to better performance. 
The proposed parameter-efficient PEQAD addresses this issue by achieving superior expressivity with a minimal number of parameters. When the complexity of the dataset challenges the expressivity of QML methods, the results demonstrate that the dynamic adjustment of trainable parameters in PEQAD, coupled with appropriate post-processing techniques, can extend the expressivity of QML algorithms. This enables them to encompass the learning target while minimizing the expressivity space. Furthermore, the low parameter requirement of our approach makes deployment on quantum hardware feasible. Under the constraint of limiting the number of trainable parameters to fewer than 1000, PEQAD consistently exhibits superior performance, often surpassing DSVDD in most scenarios. 
As demonstrated in Fig.~\ref{fig:epoch}, PEQAD attains comparable outcomes to DSVDD with a mere 200 parameters, utilising a mere 20 parameters. Moreover, augmenting the trainable parameters in PEQAD amplifies its efficacy by enhancing its expressivity. As shown in Fig.~\ref{fig:epoch}, As the number of trainable parameters in PEQAD rises from 10 to 70, its AUC exhibits an upward trend. However, this enhancement is constrained. When the parameter count becomes excessively large, leading to over-expressivity, optimisation becomes more challenging, as evidenced by the AUC dropping at the same epoch. For instance, when the parameter count reaches 240, the AUC of PEQAD performs worse than with 60 parameters.

\begin{figure*}[!htbp]
    \centering
    \subfloat[]{
        \includegraphics[width=0.4\linewidth]{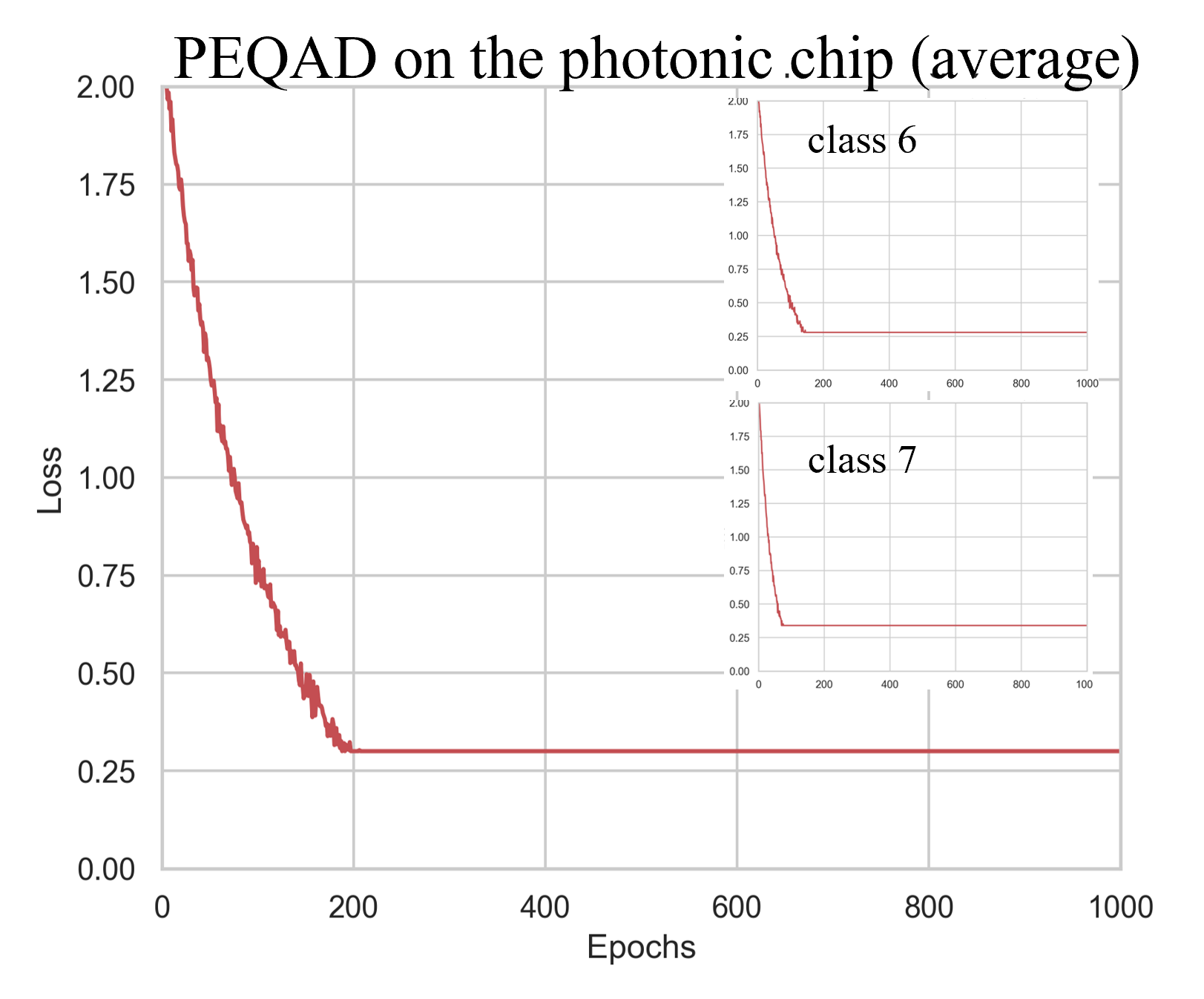}
        \label{fig:optimizationphotonic}
    }
    \hfill
    \subfloat[]{
        \includegraphics[width=0.44\linewidth]{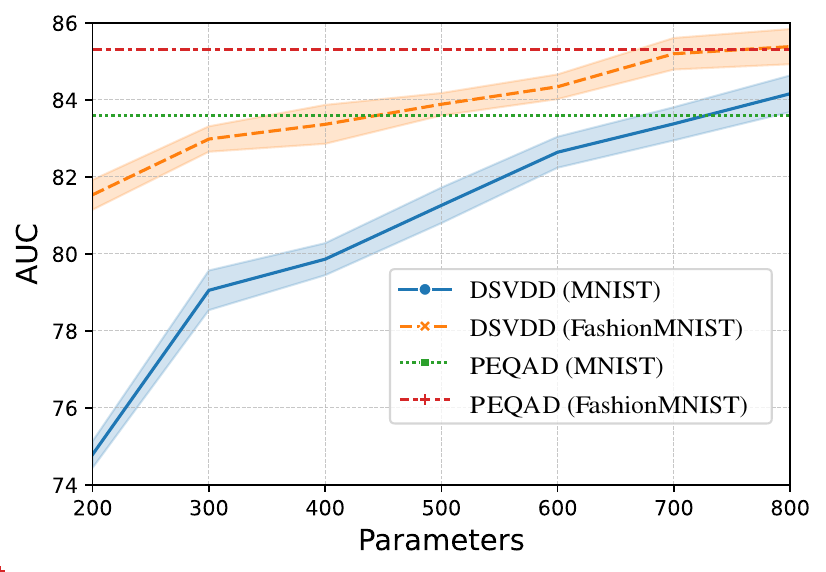}
        \label{fig:ParameterEfficiency-photonic}
    }
    \caption{(a) The optimization process of PEQAD on the photonic processor demonstrates a stable convergence with a higher final loss than that of the superconducting processor. (b) The parameter efficiency comparison between PEQAD and DSVDD on photonic quantum processors. DSVDD requires significantly more parameters to match PEQAD performance.}
    \label{fig:Parameterphotonic}
\end{figure*}

Beyond the number of parameters, PEQAD demonstrates a particular advantage in handling small datasets compared to classical deep learning algorithms. As shown in Fig.~\ref{fig:trainset}, when the number of samples in the training set is small (fewer than 400 in this case), PEQAD significantly outperforms classical DSVDD, even though DSVDD has more trainable parameters. As the training set size increases and reaches 400 samples, the performance of PEQAD stabilizes with no further improvement. On the other hand, DSVDD, with its larger number of trainable parameters, continues to improve as the training set size grows. Since the performance of PEQAD has already stabilized, we have not reported its performance beyond 800 samples, but we continue to track the performance of DSVDD.

It is worth noting that PEQAD incorporates an additional post-processing step, contributing to increased expressivity. This enhancement in expressivity is likely the reason behind PEQAD's overall superior performance on the FashionMNIST dataset compared to VQOCC. Especially on data (classes 5 and 6) that VQOCC finds difficult to handle, PEQAD still performs well. Including the post-processing step provides PEQAD with the capacity to capture intricate features and patterns in the data, resulting in enhanced anomaly detection accuracy. This result showcases the significance of careful model design in optimizing the performance of QML algorithms.

As shown in Figure \ref{fig:vqocc}, the comparison between PEQAD and VQOCC highlights that greater expressivity can enhance the performance of quantum anomaly detection algorithms. PEQAD's superior expressive capacity enables it to capture more complex patterns, translating to better performance in practical anomaly detection tasks across diverse applications. By using PEQAD to gain more powerful expressivity, researchers can further harness the power of QML to address practical anomaly detection challenges in diverse applications. Combining theoretical insights and empirical validation helps advance the understanding and practical application of PEQAD and other QML algorithms. The insights gained from this study pave the way for more robust and powerful QML techniques, bringing us closer to realizing the full potential of quantum computing in various domains.

\section{PEQAD on Other Quantum Devices}\label{Quantum Device}
PEQAD can also be implemented on other quantum hardware types, such as photonic quantum processors \cite{zhong2020jiuzhang}. We conducted preliminary experiments on a 4-qudit photonic chip, where PEQAD achieved an AUC of over 80\% using 20 parameters. Below, we present the optimization process of PEQAD on the photonic chip and compare its parameter efficiency to that of DSVDD in this setting. The optimization process of PEQAD on the photonic processor demonstrates a stable convergence with a higher final loss than that of the superconducting processor. However, due to the limited number of experimental runs, we provide these results as supplementary material. In future work, we aim to extend our validation to a broader range of quantum hardware platforms.

\section{Datasets and Code}\label{code}
All datasets used in this paper are publicly accessible: MNIST can be downloaded from \url{http://yann.lecun.com/exdb/mnist/}, Fashion-MNIST from \url{https://github.com/zalandoresearch/fashion-mnist}, and CIFAR-10 from \url{https://www.cs.toronto.edu/~kriz/cifar.html}. The code for PEQAD is available at \url{https://github.com/UCL-CCS/PEQAD.git} or \url{https://github.com/MaidaWang/PEQAD.git}.

\bibliographystyle{unsrt}
\bibliography{quantum.bib}




  

\end{document}